\documentclass[oneside,reqno,a4paper,12pt]{article}

\usepackage{graphicx}
\usepackage{amssymb,latexsym}
\usepackage{amsmath}
\renewcommand{\thefootnote}{\fnsymbol{footnote}}
\def\lsim{\hbox{ \raise.35ex\rlap{$<$}\lower.6ex\hbox{$\sim$}\ }}
\def\gsim{\hbox{ \raise.35ex\rlap{$>$}\lower.6ex\hbox{$\sim$}\ }}

\def\LEP2{{LEPII}}

\def\prl#1#2#3{    {\it Phys. Rev. Lett. }{\bf #1} (19#2) #3}

\begin{document}

\begin{center}
{\Large \bf How generic is cosmic string formation in SUSY GUTs}

\vspace{.5cm} 

Rachel Jeannerot\footnote{E-mail:rachelj@ictp.trieste.it}

\vspace{.1cm} 

{\it The Abdus Salam ICTP, strada costiera 11, 34100 Trieste, Italy }

\vspace{.1cm} 

Jonathan Rocher\footnote{E-mail:rocher@iap.fr}

\vspace{.1cm} 
\def\setC{\mathbb{C}}
\def\setR{\mathbb{R}}
{\it Institut d'Astrophysique de Paris, {${\cal G}\setR\varepsilon\setC{\cal O}$}, FRE 2435-CNRS, 98bis
boulevard Arago, 75014 Paris, France}

\vspace{.1cm} 

Mairi Sakellariadou\footnote{E-mail:msakel@cc.uoa.gr}

{\it Division of Astrophysics, Astronomy,
and Mechanics, Department of Physics,}\\ {\it University of Athens,
Panepistimiopolis, GR-15784 Zografos (Athens), Hellas}

and {\it Institut
d'Astrophysique de Paris, 98bis boulevard Arago, 75014 Paris, France}

\vspace*{0.5cm}

\end{center}                    

\begin{abstract} 
  We study cosmic string formation within supersymmetric grand unified
  theories. We consider gauge groups having a rank between 4 and 8. We examine
 all possible spontaneous symmetry breaking patterns from
  the GUT down to the standard model gauge group. Assuming standard
  hybrid inflation, we select all the models which can solve the GUT
  monopole problem, lead to baryogenesis after inflation and are
  consistent with proton lifetime measurements. We conclude that in
  all acceptable spontaneous symmetry breaking schemes, cosmic string
  formation is unavoidable. The strings which form at the end of
  inflation have a mass which is proportional to the inflationary
  scale. Sometimes, a second network of strings form at a lower
  scale. Models based on gauge groups which have rank greater than
  6 can lead to more than one inflationary era; they all end
  by cosmic string formation.
\end{abstract}

Pacs {12.10.Dm, 98.80.Cq, 11.27+d}
\renewcommand{\thefootnote}{\arabic{footnote}}

\section{Introduction}

The interface between high energy physics and cosmology is very
fruitful. Both high energy physics and cosmology enter the description
of the evolution of the early universe, at microscopic and macroscopic
levels respectively. However, cosmological models such as inflation
must originate from the particle physics model describing interactions
of the constituents of the early universe plasma.  Cosmology provides
the ground to test fundamental theories at energies far beyond the
ones accessible by any terrestrial accelerator.

The particle physics Standard Model (SM) has been tested to a very
high precision. However, evidence of neutrino
masses~\cite{SK,SNO,kamland} proves that one must go beyond this
model. The simplest explanation of the data is that neutrino get
masses via the see-saw mechanism~\cite{see-saw} which results from the
breaking of some left-right symmetry.  This is the first hint
suggesting an extension of the SM gauge group, although this is not
strictly needed since right-handed neutrinos could be present without
invoking any extra gauge symmetry. At present, Supersymmetry (SUSY) is
the only viable theory for solving the gauge hierarchy problem. In
addition, in the supersymmetric standard model the gauge coupling
constants of the strong, weak and electromagnetic interactions, with
SUSY broken at the TeV-scale, meet in a single point at around $M_{\rm
GUT} \simeq (2-3) \times 10^{16}$ GeV. This strengthens the idea that
there may be a gauge group G with a single gauge coupling constant,
which describes the interactions between particles above the scale
$M_{\rm GUT}$.  These are the so-called Supersymmetric Grand Unified
Theories (SUSY GUTs). From the point of view of cosmology, SUSY GUTs
can provide the scalar field needed for inflation, they can explain
the matter-antimatter asymmetry of the universe, and they can provide
a candidate for cold dark matter, namely the lightest superparticle
(LSP).

An acceptable SUSY GUT model should be in agreement with both the
standard model and cosmology. The grand unified gauge group must be
broken at the GUT scale down to the standard model gauge group. The
GUT gauge group must therefore contain the SM gauge group ${\rm
SU}(3)_{\rm C}\times {\rm SU}(2)_{\rm L}\times {\rm U}(1)_{\rm Y}$,
and it must predict the phenomenology that has been observed at
accelerators~\cite{PDG}.  Constraining SUSY GUTs at accelerators is a
challenge which will be undertaken in the future. On the other hand,
even if accelerators can find SUSY particles and constrain the Minimal
Supersymmetric Standard Model (MSSM), they will probably say only very
little about GUTs and symmetry breaking patterns. Luckily, a number of
new astrophysical data can be used to constrain the various schemes of
Spontaneously Symmetry Breaking (SSB) from a grand unified gauge group
down to the standard model.

In building SUSY GUTs, one faces the appearance of undesirable stable
 topological defects, mainly monopoles, but also domain walls,
 according to the Kibble mechanism~\cite{kibble}. To get rid of the
 unwanted topological defects, one employs the mechanism of
 inflation. Inflation is also the most promising mechanism for
 generating density perturbations which lead to structure formation
 and Cosmic Microwave Background (CMB) temperature anisotropies, as
 confirmed by the recent Wilkinson Microwave Anisotropy Probe (WMAP)
 measurements~\cite{wmap}.  On the other hand, inflation usually
 requires fine-tuning of its parameters, leading to the naturalness
 issue. These fine-tuning problems can be circumvented in SUSY
 models. In principle, we could build an inflationary scenario using a
 random scalar field with a given potential, which has nothing to do
 with either the SM, or a gauge theory containing the SM. This could
 come, for example, from a hidden sector. An interesting possibility
 is that inflation comes from extensions of the SM, such as a GUT
 model, which is then self-consistent: monopoles form, inflation
 originates from the GUT itself and solves the monopole problem, and
 in addition it fits with CMB data, as well as other data such as the
 baryon asymmetry which is generated by oscillations of the inflaton
 field. Models along these lines have been
 constructed~\cite{shifted1,shifted,monopb}. This scheme is the
 philosophy which we follow here.

Given an inflationary scenario, we investigate the topological defects
which may be produced at subsequent phase transitions. We consider all
possible symmetry breaking schemes and examine which kind of
topological defects are left after inflation, if any.  In all schemes,
only local topological defects can arise, since we only consider gauge
symmetries. If monopoles or domain walls are produced after inflation,
then these SSB patterns are discarded, since these defects should have
closed the universe. The only acceptable SSB patterns are those
which, after the inflationary stage, either lead to the formation of
cosmic strings or to no defects at all. If cosmic strings (topological
defects) are formed, we should examine their type (Nambu-Goto strings,
superconducting strings) and then check their compatibility
with the  constraint coming from the recent measurements of the
CMB temperature anisotropies~\cite{wmap},~\cite{bouchet}. If embedded
strings are formed, then we should examine their stability (they are in
general unstable under small perturbations). The 
symmetry breaking patterns leading to the formation of textures cannot
be constrained, since local textures decay very fast~\cite{textures}
and therefore play no role in cosmology.

We organize the rest of the paper as follows: In Section 2, we discuss
the theoretical framework of our study.  We discuss the various kinds
of topological defects which may form, and the criterion for their
formation. We briefly review the standard model for inflation in SUSY
GUTs, and we comment on leptogenesis.  In Section 3, we discuss the
choice of the gauge groups which we consider. In Section 4, we list
all possible SSB patterns from the selected GUT gauge groups down to
the standard model gauge group.  We review the most common embeddings
of the standard model in each GUT.  Each embedding leads to specific
SSB patterns. We list them all, giving the type of defect which is
formed at each phase transition.  We then discuss which of the SSB
patterns are allowed from cosmology and we count for each group the
number of schemes where strings are formed after inflation, as
compared to the number of schemes with no defect. We round up with
our conclusions in Section 5. Finally, in Appendix A, we list the
maximal sub-algebras which we employ for the groups considered in our
study.

\section{Theoretical framework}

\subsection{Topological Defects}

The assumption of a GUT implies that our universe has undergone a
series of phase transitions associated with the spontaneous symmetry
breaking of the GUT gauge group ${\rm G}_{\rm GUT}$ down to the standard
model gauge group ${\rm G}_{\rm SM} = {\rm SU}(3)_{\rm C} \times {\rm
SU}(2)_{\rm L} \times {\rm U}(1)_{\rm Y}$ at $M_{\rm GUT} \sim 3
\times 10^{16}$ GeV. The last phase transition of the SSB pattern is
the electroweak phase transition which takes place at $M_{\rm EW} \sim
10^2$ GeV as ${\rm G}_{\rm SM}$ breaks down to ${\rm SU}(3)_{\rm C} \times
{\rm U}(1)_{\rm Q}$. There might be one, more than one, or none
intermediate symmetry group between ${\rm G}_{\rm GUT}$ and ${\rm G}_{\rm
SM}$. The important cosmological consequence of these SSB schemes is
the formation of topological defects via the Kibble
mechanism~\cite{kibble}.

If we have a system with a
topologically non-trivial vacuum manifold, then fields in different
spatial regions fall into different ground states, and thus SSB may be
followed by the emergence of a network of topological defects during
the associated phase transition. (For a review on topological defects
the reader is referred to Refs.~\cite{ShelVil},~\cite{hk}). This leads
to the GUT monopole problem: all GUTs based on simple gauge groups
lead to the formation of topologically stable monopoles whose density
is about $10^{18}$ times greater than the experimental limit. Homotopy
theory tells us that topologically stable monopoles always form. Moreover,
 a wide variety of other defects may also form, leading to
important astrophysical and cosmological implications. 

In this paper, we study the formation of topological defects in
realistic GUTs considering all possible SSB patterns of a given
group. Allowing for standard hybrid inflation we can then throw away
all schemes which lead to the formation of unwanted defects and check
whether strings form at the end of inflation or after inflation has
completed.

Let us consider the symmetry breaking of a group G down to a sub-group
H of G. In order to see whether topological defects form during the
phase transition associated with the breaking of G down to H, we can
study the homotopy groups $\pi_{\rm k}({\rm G}/{\rm H})$ of the vacuum
manifold ${\cal M}_n={{\rm G}/ {\rm H}}$ . If $\pi_{\rm k}({\rm
G}/{\rm H})\neq 0$, then topological defects are formed; if $k=0$ then
domain walls form, if $k=1$ then cosmic strings form, if $k=2$ then
monopoles form, and if $k=3$ then textures appear.

Spontaneous symmetry breaking patterns which lead to the formation of
monopoles or domain walls are ruled out since they are incompatible
with our universe, except if an inflationary era took place after
their formation. The reason for which monopoles and domain walls are
undesirable, is that in both cases they soon dominate the energy
density of our universe and close it. The textures are not studied in
this work because in the local case, their relative contribution to
the energy density of the universe decreases rapidly with
time~\cite{textures}.  Thus, we cannot constrain SSB patterns with
textures because they cannot play a significant role in cosmology.

In addition to topological defects, a gauge field theory may have
non-topological defects. It is possible to obtain a sub-manifold
${\cal M}_m ~(m<n)$, of the original vacuum manifold ${\cal M}_n$, by
freezing out some combinations of the original fields. If the topology
of the ${\cal M}_m$ is such that the theory admits topological
defects, then one can create configurations of the unconstrained
fields which correspond to topological defects. Provided these
configurations satisfy the equations of motion of the unconstrained
theory, then embedded defects appear~\cite{emb1},~\cite{emb2}.  More
precisely, if we have a symmetry breaking G$\rightarrow$ H and G$_{\rm
  emb}\rightarrow$ H$_{\rm emb}$, with ${\rm G}_{\rm emb} \subset {\rm
  G}$ and ${\rm H}_{\rm emb}\subset{\rm H}$, we examine whether
$\pi_k({\rm G}_{\rm emb}/{\rm H}_{\rm emb})\neq 0$, which is the criterion for the
appearance of embedded $(2-k)$-dimensional defects.

Embedded defects are not topologically stable and in general they are
not dynamically stable either~\cite{periv}. However, a number of
mechanisms have been proposed in the literature, which may stabilize
the embedded strings and therefore, they may play an important role in
cosmology.  For example, the pion string in the theory of strong
interactions, and the electroweak Z-string in the standard electroweak
theory can be stabilized in the early universe via finite-temperature
plasma effects~\cite{stab1}. In addition, an electroweak
Z-string can be also stabilized by the presence of bound states of a
complex scalar field~\cite{stab2}. Embedded gauge monopoles always
suffer~\cite{unst} from a long range instability (the
Brandt-Neri-Coleman instability~\cite{bnc}), and therefore, we do not
consider them.

\subsection{Inflation in Supersymmetric Unified Theories}
\label{sec-susyinfl}

Inflation is at present the most appealing theory which describes the
early universe. Inflation essentially consists of a phase of
accelerated expansion which took place at a very high energy
scale. Even though only special initial conditions eventually lead to
successfully inflationary cosmologies, it has been
argued~\cite{calzsakel} that these initial conditions are precisely
the likely outcomes of quantum events occurred before the inflationary
era. Thus, inflation is itself generic~\cite{calzsakel}.  In addition,
when the principles of quantum mechanics are taken into account,
inflation provides a natural explanation for the origin of the large
scale structures and the associated temperature anisotropies in the
CMB radiation.  With the increasing data on the CMB, which seem to
confirm an early inflationary era~\cite{wmap}, one needs to find the
most natural framework for inflation which can match the
data. Inflation is most naturally realized in SUSY models. (For a
review on inflation in SUSY models the reader in referred to
Ref.~\cite{LythRiotto}.) The most natural scenario for inflation,
up-to-date, is the so-called standard hybrid inflation (the reader is
referred to Refs.~\cite{Linde},~\cite{Cop},~\cite{Dvasha}).

Let us summarize how inflation arises naturally in SUSY GUTs based on
gauge groups with rank greater or equal to five. By naturally we mean
that no extra-field, nor any extra-symmetry, are needed for inflation
except those needed to build the GUT itself.  In order to satisfy COBE
data the inflationary scale has to be $\sim 10^{15.5}$
GeV~\cite{Dvasha} which is close to the GUT scale. Note that the
problem of initial conditions is not completely solved, but the
argument is that all the fields would come out from the quantum
gravity period taking values which are of the order of the cut-off
scale of the ultimate theory, which can be taken to be the Planck
scale $M_{\rm Pl}$ or the string scale~\cite{Lindechaotic}. The
horizon problem is solved for coupling constants of the order of
$10^{-2}$. The spectral index is predicted to be very close to one (we
usually get $n \simeq 0.98$).  Supergravity (SUGRA) corrections can be
kept small~\cite{buchmullercovi}.

The superpotential for hybrid inflation in SUSY GUTs is given by:
\begin{equation}
W = \alpha S \overline{\Phi} \Phi - \mu^2 S~,
\label{eq:W1}
\end{equation}
where $S$ is a GUT singlet, $\overline{\Phi}$ and $\Phi$ are GUT Higgs
fields in complex conjugate representations which lower the rank of
the group by one unit when acquiring non-zero VEV, and $\alpha$ and
$\mu$ are two constants ($\mu$ has dimensions of mass) which can both
be taken positive with field redefinition. The superpotential given in
Eq.~(\ref{eq:W1}) is the most general superpotential consistent with an
R-symmetry under which $W \rightarrow e^{i \beta} W~, \overline{\Phi}
\rightarrow e^{-i \beta} \overline{\Phi}~, \Phi \rightarrow e^{i
  \beta} \Phi$, and $S \rightarrow e^{i \beta}$.

The potential has two minima: one valley of local minima, for $S$
greater than its critical value $S_{\rm c} = \mu /\sqrt \alpha$,
$\overline{\Phi} = \Phi$, and one global supersymmetric minimum (V=0)
at $S=0$ and $\overline{\Phi} = \Phi = \mu / \sqrt \alpha$. Imposing
chaotic initial conditions, i.e. $ S \gg S_{\rm c}$, the fields
quickly settle down the valley of local minima. The potential $V=\mu^4
\neq 0$ and inflation can take place. SUSY is broken and the one-loop
corrections to the effective scalar potential can be
calculated~\cite{Dvasha}. This gives a little tilt to the scalar
potential which helps the scalar field $S$ to slowly roll down the
valley of minima. The last 50 or so e-folds of inflation take place
much below the Planck scale. When $S$ falls below its critical value
$S_{\rm c}$, inflation stops by a waterfall regime, and the fields quickly
settle down to the global minimum of the potential and supersymmetry
is restored.  SSB occurs at the end of inflation ($\overline{\Phi}$
and $\Phi$ acquire non zero VEVs after inflation, or at most during
the last e-fold; this is GUT model dependent). This is very important
for cosmology because it implies that topological defects (if any)
form at the end of inflation with a mass per unit length $\propto \mu/
\sqrt{\alpha}$.  Henceforth no monopole nor domain walls should be
associated with the SSB induced by the $\overline{\Phi}$ and $\Phi$
VEVs. They should not form at any subsequent phase transition either.
We shall use this argument to constraint all SSB of a given ${\rm G}_{\rm
  GUT}$. It was already done for supersymmetric $\rm{SO}(10)$
models~\cite{SO(10)}. It was found that among all the SSB patterns
from $\rm{SO}(10)$ down to the standard model gauge group involving at
most one intermediate symmetry breaking scale, only three are in
agreement with observations. The proton is ``stable'' (R-parity is
conserved) and no unwanted defects form after inflation.  In all these
three SSB patterns, cosmic strings form at the end of inflation.  They
imply a mixed scenario with inflation and cosmic strings to account
for the CMB temperature anisotropies. We shall generalize this to all
GUTs predicting neutrinos masses via the see-saw mechanism.

Let us comment on CMB anisotropies from inflation and cosmic strings
in SUSY GUTs models. In these scenarios, the multipole moments $C_l$
add quadratically and they are proportional to the same scale
$\Lambda_{\rm infl} = \mu/\sqrt{\alpha}$ with a proportionality
constant which is model dependent~\cite{rjinflsusy}. This can be
rewritten as $C_l^{\rm tot} = (1-x) C_l^{\rm infl} + x C_l^{\rm str}$
where $C_l^{\rm infl}, C_l^{\rm str} \propto (\Lambda_{\rm infl} /
M_{\rm Pl})$.  Here $x$ depends on the CMB normalization for each
scenario, on the coupling constant $\alpha$ of the trilinear term
$\alpha S \overline{\Phi} \Phi$ in Eq.~(\ref{eq:W1}), on the dimension
of $\Phi$, and on the GUT itself. Local cosmic string predictions are
unfortunately not very well established in detail and range from an
almost flat spectrum~\cite{acdkss} to a single wide bump at $\ell \sim
500$~\cite{mark} with extremely rapidly decaying tail.  Recent
numerical simulations of local string networks~\cite{ringeval} confirm
the existence of a bump at around $\ell \sim 600$.  It seems that the
microphysics of the string network plays a crucial role in the height
and in the position of the bump~\cite{vhs},~\cite{wiggles}.   Studies
of mixed perturbation models (inflation + cosmic strings) impose
strong constraints on the maximum contribution of the string
network~\cite{mark, bouchet}.  The initial condition is also not taken into
account (the distribution of strings forming at the end of inflation
and their micro-structure may be very different from those concerning
strings formed at standard phase transitions). What we can conclude is
that the effect of cosmic strings on the CMB power is to lower the
height of the first acoustic peak, and to displace it to smaller
angular scales, as well as to wash out any secondary peaks~\cite{sakel2}.
In addition, topological defects induce non-Gaussian statistics, due
to their non-linear evolution~\cite{sakel2}.

So far, we have been discussing F-term inflation. D-term
inflation~\cite{d-term} requires the existence (in addition to the
GUT) of a ${\rm U}(1)$ factor with a non-vanishing Fayet-Iliopoulos term
which can only appear if $TrQ\neq 0$, where $Q$ stands for the ${\rm U}(1)$
charge~\cite{fi}. D-term inflation occurs in the following way: If
one assumes an appropriate set of discrete and continuous symmetries,
the linear term in Eq.~(\ref{eq:W1}) can be forbidden. The VEV of the fields
$\overline{\Phi}$ and $\Phi$ can be then forced to equal the
Fayet-Iliopoulos term which also sets the scale of inflation. This is
the so-called D-term inflation~\cite{bindva}. The main advantage of
D-term inflation is that it works for general Kh\"aler
potentials. However, if this extra ${\rm U}(1)$ is anomalous coming from
string theory (this would be the best way to justify its presence),
the F-term is calculated using the Green-Schwarz mechanism and would
be at the string scale which is far too high for inflation. At the end
of D-term inflation cosmic strings always form (this is easy to
understand since we are breaking a ${\rm U}(1)$ gauge symmetry). In this
case they satisfy the Bogolomny bound and their contribution to the
$C_l$'s is $x=0.75$~\cite{rjinflsusy}. The string contribution is
smaller in the F-term case~\cite{rjinflsusy} and as mentioned above,
model dependent. We conclude that D-term inflation is not consistent
with observations~\cite{bouchet}, but it does not concern us anyway,
since we are interested in GUTs based on simple gauge groups.  

Since topological defects always form at the end of standard hybrid
inflation, it is easy to conclude that at least one intermediate
symmetry breaking is needed between ${\rm G}_{\rm GUT}$ and ${\rm G}_{\rm SM}$.
One way to avoid the monopole problem in single step breaking GUTs is
to consider the first non-renormalisable term in
Eq.~(\ref{eq:W1})~\cite{shifted1}. Its effect on the scalar potential
is to ``shift'' the inflationary valley of local minima to a valley in
which the GUT Higgs have already a non-vanishing VEV, implying that
the GUT is already broken during inflation so that no topological
defects form at the end of inflation~\cite{shifted1},~\cite{shifted}.
Note that non-renormalisable terms of all orders are in general
present in the superpotential if no R-symmetry is invoked to cancel
them. However, their effect on the scalar potential is usually
negligible.  The way one can solve the monopole problem with SUSY GUTs hybrid
inflation has been discussed in Ref.~\cite{monopb}.

We thus assume standard hybrid inflation which can only occur when the
rank of the group is lowered by (at least) one unit. We can then
discuss how frequently cosmic strings form at the end of inflation
with a mass proportional to the inflationary scale as discussed above,
so that both inflation and cosmic strings contribute to the CMB
temperature anisotropies. We point out that for GUTs based on gauge
groups with rank strictly greater than five, more than one stage of
inflation can occur. This could lead to a multiple inflationary
scenario with or without cosmic strings at each stage.  If more than
one stage of the SSB pattern lowers the rank of the group, there can
be a succession of short bursts of inflation~\cite{sarkar} which occur
at different scales below the Planck scale and leave behind a
distinctive signature in the spectrum of the generated scalar density
perturbations~\cite{tetradismairi}. In our scenarios, i.e. in
the selected SSB which lead to various stages of inflation as well as
to cosmic string formation at the end of each stage, multiple
inflation combined with multiple string networks arises.

\subsection{Leptogenesis}

A cosmological scenario is incomplete if it does not discuss
baryogenesis which has to occur after inflation has taken place. 
 GUT baryogenesis is washed out by
inflation and the window left for electroweak baryogenesis is very
small. The most appealing scenario today for baryogenesis is that of
leptogenesis~\cite{Yanagida} which requires non-zero neutrino masses. This scenario is
strongly favored since the discovery of non-zero neutrino
masses~\cite{SK,SNO,kamland}. (For a review on baryogenesis scenarios
and on the cosmological arguments which they render most of them
unlikely, the reader is refered to Ref.~\cite{DineKus}.)

The most economical way for getting neutrino masses is the see-saw
mechanism~\cite{see-saw}. This requires the existence of SM gauge
singlets (the right-handed neutrinos), which must get masses around
$M_{\rm R}\sim 10^{14}$ GeV from data on neutrino
oscillations~\cite{SK,SNO,kamland}. This means, that there exist a
superpotential mass term for the right-handed neutrinos of the form
$M_{\rm R} N_i N_j$, where $i,j = 1,2,3$ and $M_{\rm R}$ is a $3 \times 3$
mass matrix. The $N_i$'s are SM singlets which couple with the MSSM
lepton doublets L and electroweak up-type Higgs $H_{\rm u}$ via the
superpotential term $h_{ij} l_i H_{\rm u} N_j$, where $h_{ij}$ is a
$3\times 3$ complex Yukawa matrix. This gives rise to a non-zero mass
matrix $M_\nu$ for the left-handed neutrinos. The basic idea of
letogenesis~\cite{Yanagida}, is that when the universe cools down and
its temperature falls below $T \sim M_{\rm R}$, the right-handed
(s)neutrinos stop being in thermal equilibrium with the surrounding
plasma and decay into (s)leptons and electroweak Higgs (higgsinos);
lepton number and CP are violated~\cite{BP}. A net lepton asymmetry is
produced which is then transformed into a baryon asymmetry via
sphaleron transitions, which are effective between $10^{12}$ and
$10^{2}$ GeV~\cite{Rubakov}. The reheating temperature in
supersymmetric models is bounded by above, i.e. $T_{\rm RH} < 10^{10}$
GeV, to avoid an overproduction of gravitinos which would overclose
the universe.

The most effective way for leptogenesis is therefore non-thermal. 
This happens for example if the inflaton field decays into
right-handed neutrinos and sneutrinos (see for example
Ref.~\cite{Lazsha}). The resulting lepton asymmetry is then
proportional to the reheating temperature, inversely proportional to the inflaton
mass, and depends on neutrino mass parameters. Constraints from
successful inflation, reheating and neutrino masses can be
satisfied. In these scenarios, the right-handed neutrino masses come
from a superpotential term $\kappa_{ij} \Phi N_i N_j$, where the GUT
Higgs field $\Phi$ is identified with the GUT Higgs field entering the
inflationary superpotential given in Eq.~(\ref{eq:W1}). This is the
same Higgs field which breaks B-L (B and L are respectively baryon
 and lepton numbers)  in GUTs, predicting right-handed
neutrinos. Such GUTs contain a ${\rm U}(1)_{\rm B-L}$ gauge symmetry and
the scale of neutrino masses is proportional to the B-L breaking
scale.

Another non-thermal process for leptogenesis is via decaying B-L
cosmic strings~\cite{leptrj}. The Higgs field responsible for string
formation is the same Higgs field which is used to break B-L. Since it
gives mass to the right-handed neutrinos, there are right-handed
neutrino zero modes trapped in B-L cosmic string cores. These are
released when cosmic string loops decay and leptogenesis takes place.
If the superpotential given in Eq.~(\ref{eq:W1}) is used for
inflation, as well as to break B-L, then B-L cosmic strings form at
the end of inflation. Such models were discussed in
Ref.~\cite{leptrj2}. In this case both processes contribute to the
lepton asymmetry of the universe: the non-thermal process from
reheating after inflation and the decay of cosmic strings.

In any case, the SSB patterns which can explain the baryon asymmetry
of the universe have the B-L gauge symmetry broken at the end or
after inflation. If inflation takes place at the B-L breaking scale,
both non-thermal scenarios will compete, somehow in the same way that
both strings and inflation can contribute to CMB anisotropies. It
would be very interesting to calculate in which proportion they
contribute to the net baryon asymmetry of the universe today.

\section{Grand Unified Theories}\label{GUTs}

GUTs can solve many of the SM problems, such as the quantization of
the electric charge, the quarks and leptons masses, and the origin of
neutrino masses. On the other hand, 
SUSY solves the gauge hierarchy problem. In the MSSM
with SUSY broken at around $10^3$ GeV the strong, weak, and
electromagnetic gauge coupling constants run with energy and reach the
same value at $M_{\rm GUT} \sim 3\times 10^{16}$ GeV. Hence SUSY GUTs
can describe particle interactions at energies above $M_{\rm GUT}$ and
it must be broken down to the standard model gauge group. In this
section, we select GUT gauge groups which lead to the correct SM
phenomenology without
fine-tuning~\cite{langacker},~\cite{ross},~\cite{slansky}.

A single value for the three gauge coupling constants of the standard
model can be obtained with a simple group or with a group which is the
direct product of $n$ identical simple groups with the addition of a
discrete symmetry ${\rm Z}_n$. Simple groups are divided into four
infinite families, ${\rm SU}(n+1),{\rm SO}(2n+1), {\rm Sp}(2n)$ and
${\rm SO}(2n)$, where $n$ denotes the rank of the group. In addition
there are five simple exceptional groups, ${\rm G}_2, {\rm F}_4, {\rm
  E}_6, {\rm E}_7, {\rm E}_8$, where the index corresponds to the rank
of the group. The basic requirement for a GUT is that it must contain
the standard model gauge group ${\rm G}_{\rm SM} = {\rm SU}(3)_{\rm C}
\times {\rm SU}(2)_{\rm L} \times {\rm U}(1)_{\rm Y}$ as a sub-group.
Its rank must therefore be greater or equal to 4, which is the rank of
${\rm G}_{\rm SM}$.  Simple groups of rank 4 are ${\rm SU}(5)$, ${\rm
  SO}(9)$, ${\rm Sp}(8)$, ${\rm SO}(8)$, ${\rm F}_4$ and we can add
the semi-simple group ${\rm SU}(3)\times {\rm SU}(3)$.  Among these
groups of rank 4, only ${\rm SU}(5)$ and ${\rm SU}(3)\times {\rm
  SU}(3)$ have complex representations, which are needed in order to
describe electroweak interactions. However, ${\rm SU}(3)\times {\rm
  SU}(3)$ cannot describe particles of integer and fractional charge
and therefore it is also excluded. Thus, the only group of rank 4
which remains is ${\rm SU}(5)$~\cite{langacker}.

In selecting GUT gauge groups, we have two additional constraints: the
group must include a complex representation which is necessary to
describe the standard model fermions, and it must be anomaly free.  In
principle, ${\rm SU}(n)$ may not be anomaly free~\cite{cmm}; more precisely
it depends on the chosen fermionic representation~\cite{langacker}. We
assume that the ${\rm SU}(n)$ groups which we use have indeed a fermionic
representation that certifies that the model is anomaly free.  With
these constraints taken into account (we do not yet require see-saw
mechanism for neutrino masses), only ${\rm SO}(4n+2)$ with $n\geq 2$,
${\rm SU}(n)$ with $n\geq 5$, and ${\rm E}_6$ can be kept. We also point out
that minimal SUSY {\rm SU}(5) is ruled out by proton lifetime measurements.

The last constraint comes from neutrino masses. The fairly recent
discovery of neutrino oscillations at Superkamiokande~\cite{SK}
implies that neutrino have a mass. The Sudbury Neutrino Observatory
(SNO)~\cite{SNO} results and the KamLAND \cite{kamland} direct
measurement of neutrino mixing have confirmed the existence of non
zero neutrino masses. Since the standard model does not predict the
existence of mass for the neutrino, we must go beyond. The simplest
possibility is to add a singlet which plays the role of right-handed
neutrino. One can also add a triplet of Higgs to the SM. But neutrino
masses are predicted in GUTs which contain a ${\rm U}(1)_{\rm B-L}$
gauge symmetry~\cite{see-saw}.  The requirement of see-saw mechanism
is our next constraint on the choice of the group. We point out that
these models can also automatically lead to R-parity
conservation~\cite{Martin} and baryogenesis via
leptogenesis~\cite{Yanagida}. SUSY GUT models that we shall select at
the end are self-consistent: they predict neutrino masses and
R-parity conservation, they solve their own monopole problem with
inflation, and at the end of inflation baryogenesis via leptogenesis
can take place.

Regarding the upper bound on the rank, we limit our study to groups
with rank, $r$, less than or equal to 8. Clearly, the choice of the
maximum rank is in principle arbitrary.  The choice of $r\leq 8$
could, in a sense, be motivated by the Horava-Witten~\cite{hw} model,
based on ${\rm E}_8\times {\rm E}_8$. Each factor ${\rm E}_8$ (rank
$r=8$) can be seen as confined in one brane. Thus, within a
four-dimensional theory (no extra dimensions), the rank can be limited
to $r=8$. To be more precise, within the framework of five consistent
string theories in ten-dimensions (i.e., type I open strings, type IIA
and IIB closed strings and the two closed heterotic strings), the rank
of the gauge group is bounded to $r\leq 22$~\cite{quevedo}. However, it
is at present believed that the five string theories are related by
strong-weak coupling dualities, and they can be seen as different
limits of one underlying theory, the M-theory. In this context, one
gets non-perturbative strings which have their own non-perturbative
gauge group, thus enhancing, by a really a lot, the maximum rank
required in perturbation theory~\cite{quevedo} (a few years ago, the
upper bound of the rank was found to be $10^5$~\cite{ub}).  Even
though we limit our study in $r\leq 8$, we believe that we still
capture the main results.  Indeed, higher rank groups lead (as one can
see in the next sections) to similar SSB patterns as the one
considered for groups of smaller rank. At last, but not least, a fully
exhaustive analysis is clearly impossible.

\section{Spontaneous Symmetry Breaking Patterns}
\label{SSB}

In the previous section, we showed that a number of constraints
 restrict the choice of symmetry groups ${\rm G}_{\rm GUT}$.  In this
 section, we study all possible spontaneous symmetry breaking patterns
 from ${\rm G}_{\rm GUT}$ down to the standard model gauge group ${\rm
 G}_{\rm SM}$ (or ${\rm G}_{\rm SM} \times Z_2$) and we look for
 defect formation. Here $Z_2$ is a sub-group of the ${\rm U}(1)_{\rm B-L}$
 gauge symmetry which is contained in various gauge groups such as
 SO(10), E(6) and ${\rm SU}(8)$. It plays the role of R-parity. Recall that
 R-parity in SUSY forbids all dimensions 3 and 4 (even dimension 5)
 baryon and lepton number violating operators, therefore forbidding
 fast proton decay. This discrete $Z_2$ symmetry can be left unbroken
 down to low energy when appropriate representations are used to
 implement the SSB pattern~\cite{Martin}. R-parity is thus an
 automatic consequence of SUSY GUTs which contain ${\rm U}(1)_{\rm
 B-L}$. Only models with unbroken $Z_2$ at low energy are consistent
 with the proton lifetime measurements. Therefore when it appears in a
 SSB scheme, we keep it unbroken down to low energy.

We only consider maximal regular sub-groups~\cite{slansky};
they are listed explicitly in Appendix~\ref{subalgebras}.  We
disregard special maximal sub-groups because it is then really non
trivial to get ${\rm G_{SM}}$ with the correct phenomenology. We write
down SSB schemes which are consistent with both group theory and
particle physics phenomenology. Some of the SSB schemes may be
extremely complicated for model building. For example, non-trivial
Higgs representations may be needed. In fact, in model building with a
minimal set of Higgs, we do not usually get many intermediate SSB
scales. Also, in going beyond one or two intermediate SSB scales, the
model loses its predictability. However, this is beyond the scope of the
systematic search we are aiming to.

For each group, there may be different ways of embedding ${\rm
G_{SM}}$ in a given maximal sub-group. We use different indices to
refer to the  embedding that we consider. The three indices
${\rm C, L, Y}$ stand for Color, Left and Hypercharge respectively,
but we use more generally the index C (respectively L) when ${\rm
SU}(n)_{\rm C} \supset {\rm SU}(3)_{\rm C}$ (respectively ${\rm
SU}(n)_{\rm L} \supset {\rm SU}(2)_{\rm L}$).  We use several other
indices which correspond to different possible embeddings of ${\rm
G_{SM}}$ in maximal sub-groups of ${\rm G}_{\rm GUT}$. They are
explained below when we list the various SSB patterns for each
group. The definition of the weak hypercharge is given where
needed. The microstructure of cosmic strings is very much
dependent on these assignments, which can imply different cosmological
and astrophysical effects such as superconductivity, non-thermal
production of baryons, lepton asymmetry or dark matter.

In order to simplify the notation, we write $4_{\rm C} ~2_{\rm L}
~2_{\rm R}$ which stands for ${\rm SU}(4)_{\rm C}\times {\rm
SU}(2)_{\rm L}\times {\rm SU}(2)_{\rm R}$, $3_{\rm C} ~2_{\rm L}
~2_{\rm R} ~1_{\rm B-L}$ for $ {\rm SU}(3)_{\rm C}\times {\rm
SU}(2)_{\rm L}\times {\rm SU}(2)_{\rm R}\times {\rm U}(1)_{\rm B-L}$,
 etc. We also use the numbers $1, 2, 2', 3$ over an arrow to
distinguish the type of gauge defect which is formed during the
corresponding phase transition: $1$ stands for monopoles, $2$ for
topological cosmic strings, $2'$ for embedded strings and $3$ for domain
walls; $0$ indicates that no defect forms. If the number is given in
brackets, it stands for the type of defect formed during the SSB of
the same gauge group down to the same sub-group $\times ~Z_2$.  As an
example, ${\rm G} \overset{1 ~(1,2)}\rightarrow {\rm H} ~(Z_2)$, means
that monopoles form when G breaks down to H, while both monopoles and
cosmic strings form when G breaks down to ${\rm H} \times Z_2$.  If
($Z_2$) appears but there is no number in brackets, it is because the
$Z_2$ appeared during a previous transition and the type of defect
which forms in the SSB with unbroken $Z_2$ is identical the one
without it.  Finally, ${\rm G} \rightarrow$ $\dots$ means that the SSB
patterns of G down to ${\rm G_{SM}}$ have already been given.

\subsection{Discrete Symmetries}
\label{sec-discrete}

We briefly discuss various discrete
symmetries which may appear during the SSB patterns. The $Z_2$ sub-group
of ${\rm U}(1)_{\rm B-L}$ which plays the role of R-parity is the only
discrete symmetry that we shall consider in the SSB patterns. It must
 be there by naturalness for keeping the proton lifetime above
the experimental limits, since we do not consider the existence of any
other symmetry than ${\rm G_{GUT}}$ at the GUT scale.

Nevertheless, we point out that some discrete $Z_n$ symmetries may be
left unbroken when the rank of the group is lowered. This depends on 
the Higgs representation which is used to implement the SSB. However,
only two discrete symmetries, the standard $Z_2$ parity and one $Z_3$
parity, are anomaly free and can remain unbroken at low
energy~\cite{ibanez}.  (Note that by adding some gauge singlets and/or
doublets two more $Z_3$ could be allowed.) Also, in order to get the
$Z_3$ symmetry some very high Higgs dimensional representations are
needed~\cite{montigny}. For example, in order to get a residual $Z_3$
from ${\rm E}_6$, one has to choose a 3003-dim Higgs representation. To
simplify our work, we disregard these $Z_n$. They must be broken at
some stage during the SSB pattern, so that they are broken today. From
a cosmological point of view, when these discrete symmetries break,
unwanted domain walls form. In a full model, they must therefore
appear and be broken before inflation.

Another discrete symmetry, $Z_2^{\rm C}$, can also be left unbroken when
Pati-Salam or Left-Right symmetry groups appear. This leads to the formation
of $Z_2$-strings which get connected via domain walls when $Z_2^{\rm
C}$ breaks~\cite{Kibblelaz}.  (This is not coming from the breaking of
a gauge ${\rm U}(1)$ symmetry and hence does not enter in the comments
above.) The $Z_2^{\rm C}$ symmetry is also known as
D-parity~\cite{moha2}. The scale of breaking of $Z_2^{\rm C}$ and
${\rm SU}(2)_{\rm R}$ may be, in principle, separated. We  discuss
all the SSB patterns for SUSY SO(10) with and without unbroken
D-parity at high scale. Although the unbroken D-parity may also
appear in ${\rm E}_6$ models, for reasons of simplification, we
do not discuss it. The important issue is that it must be broken
before inflation takes place.

\subsection{SU(5)}

The discussion of ${\rm SU}$(5) GUT is very short, since 
SU(5) has rank 4 and can only break directly down to the standard
model gauge group.  This SSB leads to the formation of topologically
stable monopoles which are inconsistent with observations. One way to
solve the monopole problem in SUSY ${\rm SU}$(5) is to introduce an extra
singlet and to give non-trivial initial conditions to the fields in the
Higgs potential~\cite{masiero}. In the following section we discuss
GUT gauge groups with rank greater or equal to five.

\subsection{SO(10)} 

SO(10) is a gauge group of rank 5, which contains as maximal
sub-groups ${\rm SU}(5) \times {\rm U}(1)$ and the Pati-Salam
gauge group ${\rm G}_{\rm PS} = {\rm SU}(4)_{\rm C}\times {\rm SU}(2)_{\rm
R}\times {\rm SU}(2)_{\rm L}$, where ${\rm SU}(4)_{\rm C} \supset
{\rm SU}(3)_{\rm C} \times {\rm U}(1)_{\rm B-L}$.

In order to give explicit definition for the hypercharge, we consider
the following decomposition~\cite{harada} 
\begin{equation}
{\rm SO}(10) \supset {\rm SU}(5) \times {\rm U}(1)_{\rm V} \supset
{\rm SU}(3)_{\rm C} \times {\rm SU}(2)_{\rm L}\times {\rm U}(1)_{\rm
Z} \times {\rm U}(1)_{\rm V}. 
\end{equation}
There are two possible assignments for the hypercharge $Y$ that
reproduce the SM and they depend on whether it is only included in
${\rm SU}(5)$ or also in {\rm SO}(10). In the first case
\begin{equation}
{Y\over 2} = Z.
\end{equation}
It is used for SSB via the Georgi-Glashow model~\cite{georgiglashow};
we add no subscript to ${\rm SU}(5)$. In the second case
\begin{equation}
{Y\over 2} = -{1\over 5} (Z+V),
\end{equation}
and it is used for the breakings via the flipped ${\rm SU}(5)$
model~\cite{flippedsu5}, in which case we add the subscript F,
i.e. we write $5_{\rm F}$.  In ${\rm SO}(10)$, all the standard model
fermions of each family plus a right-handed neutrino belong to the
16-dimensional representation. The decomposition of the 16 under ${\rm
SU}(5)$ and ${\rm SU}(5)_{\rm F}$ is given in
Refs.~\cite{georgiglashow,flippedsu5}.

Thus, there are two ways of embedding ${\rm SU}(5) \times {\rm U}(1)_{\rm V}  
\supset {\rm SU}(3)_{\rm C} \times {\rm SU}(2)_{\rm L} \times {\rm 
U}(1)_{\rm Z}$ in minimal ${\rm SO}(10)$ GUT, but there is only one way for ${\rm
SU}(2)_{\rm R}$ \cite{harada}.  Here V is related to the third 
component $I_{\rm R}^3$ of ${\rm SU}(2)_{\rm R}$ and to ${\rm B-L}$, which
is contained in ${\rm SO}(10)$ by
\begin{equation}
{\rm V} = -4 I^3_{\rm R} - 3 ({\rm B-L})~,
\end{equation}
and to $Z$ by
\begin{equation} 
{\rm Z} = -I^3_{\rm R} + {1\over 2} ({\rm B-L})~.
\end{equation}
Thus, in the first case,
\begin{equation}
{{\rm Y}\over 2} = -I_{\rm R}^3  + {1\over 2} ({\rm B-L})~,
\end{equation}
while in the second case
\begin{equation}
{{\rm Y}\over 2} = I_{\rm R}^3  + {1\over 2} ({\rm B-L})~. \label{eq:Y21}
\end{equation}

 We list below the SSB schemes of ${\rm SO}(10)$ via {\rm
SU}(5) sub-groups.  We indicate the type of defect(s) formed at each
phase transition 
\begin{equation}
\label{eq:51}
    {
  {\rm SO}(10) \left\{ \begin{array}{ccl} 
    \overset{1}{\longrightarrow} & 5   ~1_{\rm V} &
    \hspace{-.2cm}\left\{ 
    \begin{array}{cccc} 
      \overset{2~(2)}{\rightarrow} & 5 ~(Z_2) 
      &\overset{1}{\longrightarrow}&  {\rm G}_{\rm SM}~(Z_2) \\ 
      \overset{1}{\longrightarrow}& 3_{\rm C} ~2_{\rm L} ~1_{\rm Z} ~1_{\rm V} 
      &\overset{2~(2)}{\longrightarrow}& {\rm G}_{\rm SM}~(Z_2)\\ 
      \overset{1,2 ~(1,2)}{\longrightarrow}& {\rm G}_{\rm SM}~(Z_2)\\
    \end{array}
    \right.
    \\
    \overset{1}{\longrightarrow} & 5_{\rm F} ~1_{\rm V}  &
\begin{array}{cc} \overset{2'~(2)}{\longrightarrow} & {\rm G}_{\rm SM}~(Z_2)\end{array}\\
    \overset{0 ~(2)}{\longrightarrow}& 5~(Z_2) &
\begin{array}{cc}\overset{1}{\longrightarrow} & {\rm G}_{\rm SM}~(Z_2)\end{array}\\
  \end{array}
  \right.
    }
\end{equation}
    
{\rm SO}(10) can also break via the left-right symmetric groups ${\rm
G}_{\rm PS} \supset {\rm SU}(3)_{\rm C}\times {\rm SU}(2)_{\rm
R}\times {\rm SU}(2)_{\rm L}\times {\rm U} (1)_{\rm B-L}$, in which case the
assignment of the hypercharge is given by Eq.~(\ref{eq:Y21}). As
explained in the previous section, a discrete symmetry know as
D-parity (noted as $Z_2^{\rm C}$) can appear, leading to the
formation of walls bounded by strings; such configurations are not
problematic for cosmology. However, if inflation takes place before
the formation of domain walls, then these would become cosmologically
catastrophic; this situation is forbidden.  Another $Z_2$ appears when
{\rm SO}(10) is breaking via ${\rm G}_{\rm PS}$~\cite{Kibblelaz};
indeed, it is not {\rm SO}(10) but its universal covering group ${\rm
Spin}(10)$ which is really broken to $[({\rm Spin}(6)\times {\rm
Spin}(4))/Z_2](\times Z_2^{\rm C}$) (We remind to the reader that 
${\rm SU}(4) \times {\rm SU}(2) \times {\rm SU}(2)\sim {\rm
Spin}(6)\times {\rm Spin}(4)$.) The quotient $Z_2$ results from the
non-trivial intersection of ${\rm Spin}(6)$ and ${\rm Spin}(4)$ and
implies the formation of monopoles.

The SSB patterns of ${\rm G}_{\rm PS}$ and ${\rm G}_{\rm PS}$ with
D-parity down to ${\rm G_{SM}} ~(Z_2)$ are respectively given by
  \begin{equation}
\label{eq:ps}
\begin{array}{clllcccc} 
 4_{\rm C}     ~2_{\rm L}     ~2_{\rm R}   &  
\left\{ 
\begin{array}{cllllccc} 
\overset{1}{\longrightarrow} & 3_{\rm C} ~2_{\rm L} ~2_{\rm R} ~1_{\rm
B-L} & \left\{
\begin{array}{cllllccc}
 \overset{1}{\longrightarrow}  &   3_{\rm C}     ~2_{\rm L}     ~1_{\rm R}     ~1_{\rm B-L}   &  \overset{2 ~(2)}{\longrightarrow}  &   {\rm G}_{\rm SM}  ~(  Z_2  ) \\ 
  \overset{2' ~(2)}{\longrightarrow}  &   {\rm G}_{\rm SM} ~(Z_2)\\
 \end{array}
\right.
\\
  \overset{1}{\longrightarrow}  &   4_{\rm C}     ~2_{\rm L}     ~1_{\rm R}   &
\left\{ 
\begin{array}{cllllccc} 
  \overset{1}{\longrightarrow}  &   3_{\rm C}     ~2_{\rm L}     ~1_{\rm R}     ~1_{\rm B-L}   &   \overset{2 ~(2)}{\longrightarrow}  &   {\rm G}_{\rm SM} ~(Z_2)\\
 \overset{2' ~(2)}{\longrightarrow}  &   {\rm G}_{\rm SM} ~(Z_2)\\
 \end{array}
\right.
\\
  \overset{1}{\longrightarrow}  &   3_{\rm C}     ~2_{\rm L}     ~1_{\rm R}     ~1_{\rm B-L}   &  ~~~~\overset{2 ~(2)}{\longrightarrow}      ~~{\rm G}_{\rm SM} ~(Z_2)\\
  \overset{1 ~(1,2)}{\longrightarrow}  &   {\rm G}_{\rm SM}  ~(  Z_2  )\\
  \end{array}
\right.
\end{array}
\end{equation}
and
\begin{equation}  
\label{eq:psZ2}
\begin{array}{clllcccc} 
  4_{\rm C}     ~2_{\rm L}     ~2_{\rm R}     ~Z_2^{\rm C}   &  
\left\{ 
\begin{array}{cllllccc}

 \overset{1 }{\longrightarrow}  &    3_{\rm C}     ~2_{\rm L}     ~2_{\rm R}     ~1_{\rm B-L}     ~Z_2^{\rm C}   &  
\left\{ 
\begin{array}{cllllccc} 
\overset{3}{\longrightarrow}  &    3_{\rm C}     ~2_{\rm L}     ~2_{\rm R}     ~1_{\rm B-L}   &   \overset{}{\longrightarrow}   &   \cdots  \\
 \overset{1,3}{\longrightarrow}  &   3_{\rm C}     ~2_{\rm L}     ~1_{\rm R}     ~1_{\rm B-L}   &  \overset{2 ~(2)}{\longrightarrow}  &   {\rm G}_{\rm SM}  ~(  Z_2  )\\ 
  \overset{2',3 ~(2,3)}{\longrightarrow}  &   {\rm G}_{\rm SM} ~(Z_2) \\
  \end{array}
\right.
\\
  \overset{1}{\longrightarrow}  &   4_{\rm C}     ~2_{\rm L}     ~1_{\rm R}     ~Z_2^{\rm C}   &
\left\{ 
\begin{array}{cllllccc}
  \overset{3}{\longrightarrow}  &   4_{\rm C}     ~2_{\rm L}     ~1_{\rm R}   &  {\longrightarrow}  &   \cdots  \\
  \overset{1,3}{\longrightarrow}  &   3_{\rm C}     ~2_{\rm L}     ~1_{\rm R}     ~1_{\rm B-L}   &  \overset{2 ~(2)}{\longrightarrow}  &  {\rm G}_{\rm SM} ~(Z_2)\\
  \overset{3 ~(2,3)}{\longrightarrow}  &   {\rm G}_{\rm SM} ~(Z_2) \\
  \end{array}
\right.\\
\overset{3}{\longrightarrow}  &   4_{\rm C}     ~2_{\rm L}     ~2_{\rm R}     
&{\longrightarrow}   ~~~\mbox{\rm Eq. (\ref{eq:ps})}   \\
  \overset{1}{\longrightarrow}  &   4_{\rm C}     ~2_{\rm L}     ~1_{\rm R}   &  {\longrightarrow}     ~~~\cdots  \\ 
\overset{1,3}{\longrightarrow}  &   3_{\rm C}     ~2_{\rm L}     ~2_{\rm R}     ~1_{\rm B-L}   &  \overset{}{\longrightarrow}     ~~~\cdots   \\
  \overset{1, 3}{\longrightarrow}  &   3_{\rm C}     ~2_{\rm L}     ~1_{\rm R}     ~1_{\rm B-L}    &  \overset{2 ~(2)}{\longrightarrow}     ~~~ {\rm G}_{\rm SM} ~(Z_2)   \\
  \overset{1,3 ~(1,2,3)}{\longrightarrow}  &   {\rm G}_{\rm SM}  ~(  Z_2  ).\\
 \end{array}
\right. 
\end{array}
\end{equation}

The SSB schemes of {\rm SO}(10) via the left-right groups with
associated defect formation are
\begin{equation}
    {
  {\rm SO}(10) \left\{ \begin{array}{ccccc} 
  \overset{1}{\longrightarrow}  &   4_{\rm C} ~2_{\rm L}  ~2_{\rm R}   &  
\overset{}{\longrightarrow}    & \mbox{\rm Eq. (\ref{eq:ps})}
\\
      \overset{1,2}{\longrightarrow}  &   4_{\rm C} ~2_{\rm L}   ~2_{\rm R}     ~Z_2^{\rm C}   &  \overset{}{\longrightarrow}    & \mbox{{\rm Eq. (\ref{eq:psZ2})}}

\\
  \overset{1,2}{\longrightarrow}  &   4_{\rm C}     ~2_{\rm L}     ~1_{\rm R}     ~Z_2^{\rm C}   &  \overset{}{\longrightarrow}    &  \cdots  \\
  \overset{1}{\longrightarrow}  &   4_{\rm C}     ~2_{\rm L}     ~1_{\rm R}   &  \overset{}{\longrightarrow}  &    \cdots  \\
  \overset{1,2}{\longrightarrow}  &   3_{\rm C}     ~2_{\rm L}     ~2_{\rm R}     ~1_{\rm B-L}     ~Z_2^{\rm C}   &  \overset{}{\longrightarrow}  &    \cdots  \\
  \overset{1}{\longrightarrow}  &   3_{\rm C}     ~2_{\rm L}     ~2_{\rm R}     ~1_{\rm B-L}   &  \overset{}{\longrightarrow}    &  \cdots  \\
  \overset{1}{\longrightarrow}  &   3_{\rm C}     ~2_{\rm L}     ~1_{\rm R}     ~1_{\rm B-L}   &  \overset{2~(2)}{\longrightarrow}   &   {\rm G}_{\rm SM}  ~(  Z_2  )\\
  \overset{1 ~(1,2)}{\longrightarrow}  &   {\rm G}_{\rm SM}  ~(Z_2)
\\
  \end{array}
  \right.
    }
\end{equation}

 The SSB patterns listed above and the type of defect indicated
above the arrows, contain all the information one needs to address the
question of whether cosmic strings (topological or embedded) are
expected to exist in models which are compatible with both particle
physics and cosmology. The acceptable models must be consistent with
proton lifetime measurements, solve the GUT monopole problem with
inflation, and explain the baryon asymmetry of the universe. Inflation
takes place when the rank of the group is lowered and non-thermal
leptogenesis (i.e.  B-L breaks at the end of inflation) is efficient.
For GUTs based on gauge groups which have rank less or equal to five
such as SO(10) or SU(6), in each SSB pattern, there is one single
choice for the phase transition where hybrid inflation can take place.
In SO(10), non-thermal leptogenesis always takes place at the end of
inflation.  On the other hand, for GUTs based on gauge groups which
have rank greater than five, there may be more than one choice for the
phase transition which leads to inflation. In these GUTs where
inflation can take place at different stages in the SSB patterns, B-L
is not necessarily broken at the end of inflation. Models satisfying
all constraints must lead to efficient leptogenesis; B-L must be
broken at the end of inflation.

Since in standard hybrid inflation SSB takes place at the end of
inflation, in the schemes which are consistent with cosmology from a
defect point of view, inflation can only take place during a given
phase transition, with no monopoles or domain walls at this or at a subsequent
phase transition.

 For SO(10), we find that there are 68 SSB patterns which do not lead
to  formation of unwanted defects after inflation and all these models
lead to the formation of cosmic strings (topological cosmic strings or
embedded ones) at the end of inflation. More precisely, we find that
there are 34 SSB patterns with cosmic strings (topological defects)
and unbroken matter parity, i.e. ${\rm G}_{\rm SM}\times Z_2$. There
are 21 SSB patterns leading to cosmic string formation, but with
broken R-parity. Finally, there are 13 SSB schemes with embedded
strings. In SO(10), when embedded strings are formed, R-parity is
always broken. In all these models, B-L is broken at the end of
inflation and leptogenesis is efficient. As discussed earlier, the
proton lifetime measurements require unbroken R-parity. There are
therefore only 34 SSB patterns which satisfy all the constraints and
they all lead to the formation of topological cosmic strings at the
end of inflation.

\subsection{SU(6)}  

{\rm SU}(6) is the second group of rank 5. The maximal sub-groups of
{\rm SU}(6) are given in Table~\ref{sub-group1}. Recall that {\rm
SU}(6) does not contain B-L and, therefore it cannot accommodate the
data on neutrino oscillations. There are only few possibilities for
the spontaneous symmetry breaking patterns from {\rm SU}(6) down to
the ${\rm G}_{\rm SM}$. We list them below, indicating also the type
of defects, if any, formed at each phase transition.
\begin{equation}
    {
{\rm SU}(6) \left\{ \begin{array}{cllll} 
   \overset{1}{\longrightarrow}& 5 ~1_6 & 
   \left\{ 
    \begin{array}{llll}
\overset{2}{\longrightarrow}& 5 & \overset{1}{\longrightarrow} & {\rm G}_{\rm SM} \\
\overset{1}{\longrightarrow}& 3_{\rm C} ~2_{\rm L} ~1 ~1 & \overset{2}{\longrightarrow} & {\rm G}_{\rm SM}\\
\overset{1,2}{\longrightarrow}& {\rm G}_{\rm SM} \\
\end{array}\right.
\\
\overset{1}{\longrightarrow}& 4_{\rm C} ~2_{\rm L} ~1 
&\left\{ \begin{array}{llll} 
\overset{1}{\longrightarrow}& 3_{\rm C} ~2_{\rm L}~1 ~1 &\overset{2}{\longrightarrow} & {\rm G}_{\rm SM} \\
\overset{2'}{\longrightarrow}& {\rm G}_{\rm SM}\\
\end{array}\right.
\\
\overset{1}{\longrightarrow}& 3_{\rm C} ~3_{\rm L}~1 
&\left\{ \begin{array}{llll} 
\overset{1}{\longrightarrow}& 3_{\rm C} ~2_{\rm L}~1 ~1 & \overset{2}{\longrightarrow} & {\rm G}_{\rm SM} \\
\overset{2'}{\longrightarrow} & {\rm G}_{\rm SM}\\
\end{array}\right.
\\
\overset{1}{\longrightarrow}& 3_{\rm C} ~2_{\rm L}~1 ~1 & \overset{2}{\longrightarrow} ~~~{\rm G}_{\rm SM} \\
\overset{0}{\longrightarrow}& 5 & \overset{1}{\longrightarrow} ~~~{\rm G}_{\rm SM} \\
\overset{1}{\longrightarrow} & {\rm G}_{\rm SM}\\
  \end{array}
  \right.
    }
\end{equation}
  
Following the same approach as in the case of SO(10), one finds that
there are six cosmologically allowed SSB schemes (from a defect point
of view).  They all lead to the formation of cosmic strings at the end
of inflation. There are four schemes with topological strings and two
with embedded ones. However, SU(6) does not contain ${\rm U}(1)_{\rm
B-L}$, data on neutrino oscillations cannot be accommodated without
extension of the minimal version and R-parity is not there. Thus,
minimal SU(6) is not an acceptable group for particle physics.

\subsection{${\rm \bf E}_6$}  

${\rm E}_6$ is a group of rank 6 and it has three regular maximal
sub-groups which can accommodate the standard model
$${\rm SO}(10) \times {\rm U}(1)_{\rm V'}$$
$${\rm SU}(3)_{\rm C}\times {\rm SU}(3)_{\rm L}\times {\rm SU}(3)_{\rm (R)}$$
$${\rm SU}(6)\times {\rm SU}(2)$$ We study the SSB patterns of ${\rm
E}_6$ via each of them in the following sections.  (We follow the notation of
Ref.~\cite{harada}).

\subsubsection{Breaking ${\rm \bf E}_6$ via ${\rm SO}(10) \times {\rm U}(1)$}

Let us start with ${\rm E}_6 \supset {\rm SO}(10) \times {\rm
U}(1)_{\rm V'}$ and ${\rm SO}(10) \supset {\rm SU}(5)\times {\rm
U}(1)_{\rm V}$.  There are three possible assignments for the
hypercharge Y which reproduce the SM depending on whether ${\rm
U}(1)_{\rm Y} \subset {\rm SU}(5)$, or ${\rm U}(1)_{\rm Y} \subset
{\rm SO}(10)$, or ${\rm U}(1)_{\rm Y} \subset {\rm E}_6$. They are
respectively given by~\cite{harada}
\begin{equation}
{Y\over 2} = Z \label{eq:Y1}
\end{equation}
\begin{equation}
{Y\over 2} =-{1\over 5} (Z+V) \label{eq:Y2}
\end{equation}
~\begin{equation}
{Y\over 2} =-{1\over 20} (4Z-V-5V') \label{eq:Y3}
\end{equation}
So the ${\rm U}(1)_{\rm Y}$ with hypercharge given in Eq.~(\ref{eq:Y1}) is
only contained in {\rm SU}(5) and is valid for the breakings through
the Georgi-Glashow model. The hypercharge in Eq.~(\ref{eq:Y2}) is
contained in SO(10) and is the one appearing in the breakings through
flipped ${\rm SU}(5)$. Finally the last assignment of $Y$ in
Eq.~(\ref{eq:Y3}) correspond to ${\rm U}(1)_{\rm Y} \subset {\rm E}_6$. This
is the sub-group which appears in the breaking of  ${\rm E}_6$ through the
E-twisted ${\rm SU}(5)$ model for example.  We distinguish the ${\rm
SU}(5)$ of each of these three cases by writing it as $5$, $5_{\rm F}$,
or $5_{\rm E}$ respectively. The SSB patterns for $5 ~1_{\rm V}
~1_{\rm V'}$ and $5_{\rm F} ~1_{\rm V} ~1_{\rm V'}$ are respectively
given by
\begin{equation}
\label{eq:5}
\begin{array}{clllllccc}
 5  ~1_{\rm V}   ~1_{\rm V'}  
& \left\{ 
\begin{array}{clllllccc}
\overset{2 ~(2)}{\longrightarrow} & 5  ~1_{\rm V'}  ~( Z_2 ) 
& \left\{ 
\begin{array}{clllllccc}
 \overset{2}{\longrightarrow} & 5 ~( Z_2 ) & \overset{1}{\longrightarrow} &  {\rm G}_{\rm SM}  ~( Z_2 ) \\
 \overset{1}{\longrightarrow} &  {\rm G}_{\rm SM}   ~1_{\rm V'}  ~( Z_2 ) & \overset{2}{\longrightarrow} &  {\rm G}_{\rm SM}  ~( Z_2 )\\
 \overset{1,2}{\longrightarrow} &  {\rm G}_{\rm SM}  ~( Z_2 )\\
\end{array}
  \right.
\\
\overset{1}{\longrightarrow} &  {\rm G}_{\rm SM}   ~1_{\rm V}   ~1_{\rm V'}   
& \left\{ 
\begin{array}{cllllllccc}
\overset{2}{\longrightarrow} &  {\rm G}_{\rm SM}   ~1_{\rm V}  & \overset{2~(2)}{\longrightarrow} &  {\rm G}_{\rm SM}  ~( Z_2 ) \\
 \overset{2~(2)}{\longrightarrow} &  {\rm G}_{\rm SM}   ~1_{\rm V'}  ~( Z_2 ) & \overset{2}{\longrightarrow} &  {\rm G}_{\rm SM}  ~( Z_2 )\\
\end{array}
  \right.
\\
 \overset{2}{\longrightarrow} & 5  ~1_{\rm V} & \overset{}{\longrightarrow}   ~~~~\cdots  \\
 \overset{1,2 ~(1,2)}{\longrightarrow} &  {\rm G}_{\rm SM}  ~(Z_2)\\
\end{array}
  \right.\end{array}
\end{equation}
where the hypercharge is given by Eq.~(\ref{eq:Y1}), and
\begin{equation}
\label{eq:5F}
\begin{array}{clllllccc}
 5_{\rm F}   ~1_{\rm V}   ~1_{\rm V'}   
& \left\{ 
\begin{array}{cllllccc}
~\overset{2}{\longrightarrow} &  ~~5_{\rm F}   ~1_{\rm V}   & \overset{2'~(2)}{\longrightarrow}   & {\rm G}_{\rm SM} ~(Z_2) \\
 ~\overset{2, 2'~(2)}{\longrightarrow} &  ~{\rm G}_{\rm SM} ~(Z_2) \\
\end{array}
  \right.
\end{array}
\end{equation}
where the hypercharge is given by Eq.~(\ref{eq:Y2}). In the E-twisted case the
hypercharge is given by Eq.~(\ref{eq:Y3}) and $5_{\rm E} ~1_{\rm V}
~1_{\rm V'}$ can only break down to ${\rm G}_{\rm SM} ~(Z_2)$. 

The SSB patterns with ${\rm E}_6 \supset {\rm SO}(10) \times
{\rm U}(1)_{\rm V'} \supset {\rm SU}(5)\times {\rm U}(1)_{\rm V}\times
{\rm U}(1)_{\rm V'}$ are therefore given by
\begin{equation}
    {\begin{array}{cllllll}
{\rm E}_6
\overset{1}{\rightarrow} 10 ~1_{\rm V'}  & 
  \left\{ 
    \begin{array}{cllllccccc}
 \overset{2}{\longrightarrow}  & 10  & \overset{}{\longrightarrow} ~~~~~\cdots \\
 \overset{1}{\longrightarrow} & 5  ~1_{\rm V}   ~1_{\rm V'}  & \overset{}{\longrightarrow} ~~~ \mbox{{\rm Eq.~(\ref{eq:5})}} \\
 \overset{1}{\longrightarrow}  &  5_{\rm F}   ~1_{\rm V}   ~1_{\rm V'}   
& \overset{}{\longrightarrow} ~~~ \mbox{{\rm Eq.~(\ref{eq:5F})}}
\\
 \overset{1}{\longrightarrow}  &  5_{\rm E}   ~1_{\rm V}   ~1_{\rm V'}  & \overset{2'~(2)}{\longrightarrow}   ~~{\rm G}_{\rm SM}  ~( Z_2 )\\
 \overset{0 ~(2)}{\longrightarrow} & 5  ~1_{\rm V'}  ~( Z_2 )  & \overset{}{\longrightarrow}   ~~~~\cdots \\
 \overset{1,2}{\longrightarrow} & 5  ~1_{\rm V}  & \overset{}{\longrightarrow}  ~~~~~\cdots \\
 \overset{2 ~(2)}{\longrightarrow} & 5 ~( Z_2 ) & \overset{1}{\longrightarrow}  ~~{\rm G}_{\rm SM}  ~( Z_2 )\\
  \overset{1}{\longrightarrow}  &  5_{\rm F}   ~1_{\rm V} & \overset{2' ~(2)}{\longrightarrow}  ~~{\rm G}_{\rm SM}  ~( Z_2 )\\
\overset{1}{\longrightarrow} &  {\rm G}_{\rm SM}   ~1_{\rm V}  & \overset{2 ~(2)}{\longrightarrow}   ~~{\rm G}_{\rm SM}  ~( Z_2 )\\
 \overset{1 ~(1,2)}{\longrightarrow} &  {\rm G}_{\rm SM}   ~1_{\rm V'}  ~( Z_2 ) & \overset{2}{\longrightarrow}   ~~{\rm G}_{\rm SM}  ~( Z_2 )\\
 \overset{1}{\longrightarrow} &  {\rm G}_{\rm SM}  ~( Z_2 )  
\\
  \end{array}
  \right.
    \end{array}  }
\end{equation}

${\rm SO}(10)$ in ${\rm E}_6 \supset {\rm SO}(10) \times {\rm U}(1)_{\rm V'}$
can also break via ${\rm SU}(4) \times {\rm SU}(2) \times {\rm SU}(2)$
which can be the Pati-Salam group ${\rm G}_{\rm PS}$, or ${\rm SU}(4)_{\rm
C'} \times {\rm SU}(2)_{\rm L} \times {\rm SU}(2)_{\rm G}$~\cite{sato}.  In
the first case, ${\rm U}(1)_{\rm V'}$ is orthogonal to ${\rm G_{SM}}$,
and the hypercharge assignment is exactly the same as in the ${\rm
SO}(10)$ case (Eq.~(\ref{eq:Y21})). In the second case, the
hypercharge is given by
\begin{equation}
\label{eq:c'}
Y=\frac{1}{4}V'-\frac{1}{12}C'
\end{equation} 
where $C'$ is the fifteenth generator of ${\rm SU}(4)_{\rm
C'}$. 

The SSB patterns with ${\rm E}_6 \supset {\rm SO}(10) \times
{\rm U}(1)_{\rm V'} \supset {\rm SU}(4)\times {\rm SU}(2) \times {\rm
SU}(2) \times {\rm U}(1)_{\rm V'}$ with associated defects are
\begin{equation}
    {\begin{array}{cllll} {\rm E}_6 \overset{1}{\rightarrow} 10
~1_{\rm V'} & \left\{
    \begin{array}{cllllccccc}
\overset{1,2}{\longrightarrow}& 4_{\rm C} ~2_{\rm L} ~2_{\rm R} ~1_{\rm V'} 
& \overset{}{\longrightarrow} ~~~ \mbox{{\rm Eq.~(\ref{eq:ps6})}}
\\
\overset{0}{\longrightarrow}& 4_{\rm C'} ~2_{\rm L} ~2_{\rm G} ~1_{\rm V'} 
& \overset{}{\longrightarrow} ~~~ \mbox{{\rm Eq.~(\ref{eq:psp6})}}
\\
\overset{1}{\longrightarrow}& 4_{\rm C} ~2_{\rm L} ~2_{\rm R} & \overset{}{\longrightarrow} ~~~\mbox{{\rm Eq. (\ref{eq:ps})}}\\
\overset{1}{\longrightarrow}& 3_{\rm C} ~2_{\rm L} ~2_{\rm R} ~1_{\rm B-L} ~1_{\rm V'} & \overset{}{\longrightarrow} ~~~~\cdots\\
\overset{1}{\longrightarrow}& 3_{\rm C} ~2_{\rm L} ~1_{\rm R} ~1_{\rm B-L} ~1_{\rm V'} & \overset{}{\longrightarrow} ~~~~\cdots 
\\
  \end{array}
  \right.\end{array}
    }
\end{equation}
where
\begin{equation}
\label{eq:ps6}
\begin{array}{clllllccc}
4_{\rm C} ~2_{\rm L} ~2_{\rm R} ~1_{\rm V'} 
&\left\{ 
\begin{array}{cllllccc}
\overset{2}{\longrightarrow}& 4_{\rm C} ~2_{\rm L} ~2_{\rm R} & \overset{}{\longrightarrow} ~~~\mbox{{\rm Eq.~(\ref{eq:ps})}} \\
\overset{1}{\longrightarrow}& 3_{\rm C} ~2_{\rm L} ~1_{\rm R} ~1_{\rm B-L} ~1_{\rm V'} 
& \left\{ 
\begin{array}{cllllllccc}
\overset{2}{\longrightarrow}& 3_{\rm C}  ~2_{\rm L} ~1_{\rm R} ~1_{\rm B-L} &\overset{2 ~(2)}{\longrightarrow}&  {\rm G}_{\rm SM}~(Z_2)\\
\overset{2 ~(2)}{\longrightarrow}& {\rm G}_{\rm SM} 1_{\rm V'}~(Z_2) &\overset{2}{\longrightarrow}& {\rm G}_{\rm SM}~(Z_2)\\
\overset{2 ~(2)}{\longrightarrow}& {\rm G}_{\rm SM}~(Z_2)\\
  \end{array}
  \right.
\\
\overset{1}{\longrightarrow}& 3_{\rm C} ~2_{\rm L} ~2_{\rm R} ~1_{\rm B-L} ~1_{\rm V'} 
& \left\{ 
\begin{array}{cllllccc}
\overset{1}{\longrightarrow}& 3_{\rm C} ~2_{\rm L} ~1_{\rm R} ~1_{\rm B-L} ~1_{\rm V'} &\overset{}{\longrightarrow}& \cdots\\
\overset{2}{\longrightarrow}& 3_{\rm C} ~2_{\rm L} ~2_{\rm R} ~1_{\rm B-L}&\overset{}{\longrightarrow}& \cdots\\
\overset{2' ~(2)}{\longrightarrow}& {\rm G}_{\rm SM} 1_{\rm V'}~(Z_2) &\overset{2}{\longrightarrow}& {\rm G}_{\rm SM}~(Z_2)\\
\overset{1,2}{\longrightarrow}& 3_{\rm C} ~2_{\rm L} ~1_{\rm R} ~1_{\rm B-L} &\overset{2 ~(2)}{\longrightarrow}&  {\rm G}_{\rm SM}~(Z_2)\\
  \end{array}
  \right.
\\
\overset{1}{\longrightarrow}& 4_{\rm C} ~2_{\rm L} ~1_{\rm R} 1_{\rm V'}
& \left\{ 
\begin{array}{cllllccc}
 \overset{2}{\longrightarrow}& 4_{\rm C} ~2_{\rm L} ~1_{\rm R}  &\overset{}{\longrightarrow}& \cdots\\
\overset{1}{\longrightarrow}& 3_{\rm C} ~2_{\rm L} ~1_{\rm R} ~1_{\rm B-L} ~1_{\rm V'}  &\overset{}{\longrightarrow}& \cdots\\
\overset{2' ~(2)}{\longrightarrow}& {\rm G}_{\rm SM} 1_{\rm V'}~(Z_2) &\overset{2}{\longrightarrow}& {\rm G}_{\rm SM}~(Z_2)\\
\overset{1,2}{\longrightarrow}& 3_{\rm C} ~2_{\rm L} ~1_{\rm R} ~1_{\rm B-L} &\overset{2 ~(2)}{\longrightarrow}& {\rm G}_{\rm SM}~(Z_2) \\
\overset{2 ~(2)}{\longrightarrow}& {\rm G}_{\rm SM}~(Z_2)\\
 \end{array}
  \right.
\\
\overset{1 ~(1,2)}{\longrightarrow}& {\rm G}_{\rm SM} 1_{\rm V'}~(Z_2) & \overset{2}{\longrightarrow} ~~~{\rm G}_{\rm SM}~(Z_2)\\
\overset{1,2}{\longrightarrow}& 3_{\rm C} ~2_{\rm L} ~2_{\rm R} ~1_{\rm B-L} & \overset{}{\longrightarrow} ~~~\cdots\\
\overset{1}{\longrightarrow}& 3_{\rm C} ~2_{\rm L} ~1_{\rm R} ~1_{\rm B-L} & \overset{2 ~(2)}{\longrightarrow} ~~~{\rm G}_{\rm SM}~(Z_2)\\
\overset{1,2 ~(1,2)}{\longrightarrow}& {\rm G}_{\rm SM}~(Z_2)\\
  \end{array}
  \right.\end{array}
\end{equation}
and
\begin{equation}
\label{eq:psp6}
\begin{array}{clllllccc}
4_{\rm C'} ~2_{\rm L} ~2_{\rm G} ~1_{\rm V'}
& 
\left\{ 
\begin{array}{cllllccc}
\overset{1}{\longrightarrow}& 4_{\rm C'} ~2_{\rm L} ~1_{\rm G} ~1_{\rm V'} 
\left\{ 
\begin{array}{cllllccc}
\overset{2}{\longrightarrow}& 4_{\rm C'} ~2_{\rm L} ~1_{\rm V'} &\overset{2'}{\longrightarrow}& {\rm G}_{\rm SM}\\ 
\overset{2'}{\longrightarrow}& {\rm G}_{\rm SM}\\
  \end{array}
  \right.
\\
\overset{0}{\longrightarrow}& 4_{\rm C'} ~2_{\rm L} ~1_{\rm V'} ~~~~\overset{2'}{\longrightarrow} ~~~~{\rm G}_{\rm SM}\\
\overset{2'}{\longrightarrow}& {\rm G}_{\rm SM}\\
 \end{array}
  \right.
\end{array}
\end{equation}
\vspace{.5cm}

There are also more direct schemes with these embeddings: 
\begin{equation}
    {
E_6\left\{ \begin{array}{clllllcccc} 
&\overset{1}{\longrightarrow}& 5 ~1_{\rm V} ~1_{\rm V'} &\overset{}{\longrightarrow}& \mbox{{\rm Eq.(\ref{eq:5})}}\\
&\overset{1}{\longrightarrow}& 5_{\rm F} ~1_{\rm V} ~1_{\rm V'} &\overset{}{\longrightarrow}& \mbox{{\rm Eq.(\ref{eq:5F})}}\\
&\overset{1}{\longrightarrow}& 5_{\rm E} ~1_{\rm V} ~1_{\rm V'} &\overset{2' ~(2)}{\longrightarrow}& {\rm G}_{\rm SM}~(Z_2)\\
&\overset{1}{\longrightarrow}& 5 ~1_{\rm V}  &\overset{}{\longrightarrow}& \cdots\\
&\overset{1}{\longrightarrow}& 5 ~1_{\rm V'}  &\overset{}{\longrightarrow}& \cdots\\
&\overset{0}{\longrightarrow}& 5  &\overset{1}{\longrightarrow}& {\rm G}_{\rm SM}\\
&\overset{1}{\longrightarrow}& 5_{\rm F} ~1_{\rm V}  &\overset{2'~(2) }{\longrightarrow}& {\rm G}_{\rm SM}~(Z_2)\\
&\overset{1}{\longrightarrow}& 4_{\rm C} ~2_{\rm L} ~2_{\rm R}
~1_{\rm V'} &\overset{}{\longrightarrow}& \mbox{\rm Eq. (\ref{eq:ps6})}\\

&\overset{1}{\longrightarrow}& 4_{\rm C'} ~2_{\rm L} ~2_{\rm G}
~1_{\rm V'} &\overset{}{\longrightarrow}& \mbox{\rm Eq. (\ref{eq:psp6})}\\
&\overset{1}{\longrightarrow}& 4_{\rm C} ~2_{\rm L} ~2_{\rm R}
&\overset{}{\longrightarrow}& \mbox{{\rm Eq.~(\ref{eq:ps})}}\\ 
&\overset{1}{\longrightarrow}& 4_{\rm C} ~2_{\rm L} ~1_{\rm R}&\overset{}{\longrightarrow}& \cdots\\
&\overset{1}{\longrightarrow}&
4_{\rm C} ~2_{\rm L} ~1_{\rm R} ~1_{\rm V'}
&\overset{}{\longrightarrow}& \cdots\\ &\overset{1}{\longrightarrow}&
4_{\rm C'} ~2_{\rm L} ~1_{\rm V'} &\overset{2'}{\longrightarrow}&
{\rm G}_{\rm SM}\\ &\overset{1}{\longrightarrow}& 3_{\rm C} ~2_{\rm L}
~2_{\rm R} ~1_{\rm B-L} ~1_{\rm V'} &\overset{}{\longrightarrow}&
\cdots\\ 
&\overset{1}{\longrightarrow}& 3_{\rm C} ~2_{\rm L} ~1_{\rm R} ~1_{\rm
B-L} ~1_{\rm V'} &\overset{}{\longrightarrow}& \cdots\\
&\overset{1}{\longrightarrow}& 3_{\rm C} ~2_{\rm L} ~1_{\rm R} ~1_{\rm
B-L} &\overset{2 ~(2)}{\longrightarrow}& {\rm G}_{\rm SM}~(Z_2)\\
&\overset{1}{\longrightarrow}& {\rm G}_{\rm SM} ~1_{\rm V}  &\overset{2 ~(2)}{\longrightarrow}& {\rm G}_{\rm SM}~(Z_2) \\
&\overset{1 ~(1,2)}{\longrightarrow}& {\rm G}_{\rm SM} ~1_{\rm V'} ~(Z_2) &\overset{2}{\longrightarrow}& {\rm G}_{\rm SM}~(Z_2) \\
&\overset{1 ~(1,2)}{\longrightarrow}& {\rm G}_{\rm SM}~(Z_2) \\
  \end{array}
  \right.
    }
\end{equation}

E(6) is a group of rank six, and therefore there are a priori
two possible choices for the onset of inflation in each SSB
pattern. We consider first the SSB schemes which are compatible with
observations from a defect point of view and then we add the
constraint coming from leptogenesis. We recall that for non-thermal
leptogenesis, B-L must break at the end of inflation. Our results are
the following: we find that there are in total 382 SSB patterns which
are consistent from a defect point of view,  184 leading to topological
strings and conserved R-parity, 146 with topological strings and broken
R-parity and 51 with embedded strings. There is 1 SSB scheme with no
defect fomation after the inflationary era; however R-parity is
broken. Once non-thermal leptogenesis constraint is included, there
remain 146 schemes with topological strings and conserved R-parity,
101 with topological strings and broken R-parity and 51 with embedded
strings. The total number of SSB patterns which satisfy all constraints
is 146. All of them lead to the formation of topologically stable
cosmic strings whose mass per unit length can be computed and is
proportional to the inflationary scale.

\subsubsection{Breaking ${\rm E}_6$ via ${\rm SU}(3)_{\rm C} \times
 {\rm SU}(3)_{\rm L} \times {\rm SU}(3)_{\rm R}$}

We proceed with ${\rm E}_6 \supset {\rm SU}(3)_{\rm C} \times {\rm
  SU}(3)_{\rm L}\times {\rm SU}(3)_{\rm R} \supset {\rm SU}(3)_{\rm C}
\times {\rm SU}(2)_{\rm L}\times {\rm U}(1)_{\rm Y_L} \times {\rm
  SU}(3)_{\rm R}$. There exist three different ${\rm SU}(2)$
sub-groups of ${\rm SU}(3)_{\rm R}$, namely ${\rm SU}(3)_{\rm
  R}\supset {\rm SU}(2)_{\rm R}\times {\rm U}(1)_{\rm Y_{R}}$, ${\rm
  SU}(3)_{\rm R}\supset {\rm SU}(2)'_{\rm R}\times {\rm U}(1)_{\rm
  Y'_{R}}$ and ${\rm SU}(3)_{\rm R}\supset {\rm SU}(2)_{\rm E}\times
{\rm U}(1)_{\rm Y_{E}}$. Following Ref.~\cite{harada}, we use the
notation ${\rm SU}(2)_{\rm (R)}$ which can stand for any of the the
three groups ${\rm SU}(2)_{\rm R}$, ${\rm SU}(2)'_{\rm R}$, or ${\rm
  SU}(2)_{\rm E}$. Identically, $ {\rm U}(1)_{\rm Y_{(R)}}$ can stand
for ${\rm U}(1)_{\rm Y_{R}}$, or ${\rm U}(1)_{\rm Y'_{R}}$ or ${\rm
  U}(1)_{\rm Y_{R}}$.

There are again three possible assignments for the hypercharge which
are given in Eqs.~(\ref{eq:Y1}), (\ref{eq:Y2}) and (\ref{eq:Y3}). They
can also be expressed in terms of $I_{\rm (R)}^3$, the third component
of ${\rm SU}(2)_{({\rm R})}$, and $Y_{\rm L}$, and $Y_{\rm (R)}$, the
quantum numbers of ${\rm U}(1)_{\rm Y_L}$ and ${\rm U}(1)_{\rm
Y_{(R)}}$ (we refer the reader to Ref.~\cite{harada} for details).

If ${\rm U}(1)_{\rm B-L}\subset {\rm E}_6$ is imposed, there are also three
possible assignments of B-L~\cite{harada}:

\begin{equation}
{\rm B-L} = -{1\over 5} (V-4Z) = {2\over3} Y_{\rm L} + {2\over3} Y_{\rm R}~,
\label{eq:B-L1}
\end{equation}
or
\begin{equation}
{\rm B-L} = {1\over 20} (16 Z + V + 5 V') = {2\over3} Y_{\rm L} + {2\over3} Y'_{\rm R}~,
\label{eq:B-L2}
\end{equation}
or
\begin{equation}
{\rm B-L} = -{1\over 20} (8 Z + 3 V - 5 V') = {2\over3} Y_L + {2\over3} Y_{\rm E}~.
 \label{eq:B-L3}
\end{equation}

It was shown in Ref.~\cite{harada} that among these $3 \times 3 = 9$
possible assignements for the hypercharge Y and B-L only 6 are
consistent with the standard model because ${\rm U}(1)_{\rm B-L}$ and
${\rm U}(1)_{\rm Y}$ cannot be orthogonal to the same ${\rm
SU}(2)_{({\rm R})}$ sub-group of ${\rm SU}(3)_{\rm R}$.  The possible
relations between Y, and B-L can be expressed in terms of $I_{({\rm R})}^3$. We
have the following 6 possibilities:
\begin{equation}
{Y\over 2} = -I^{3}_{\rm R} + {1\over 2} ({\rm B-L}) = -I^{3'}_{\rm R} + {1\over
2} ({\rm B-L})~,
\end{equation}
where ${\rm B-L}$ is respectively given by Eqs.~(\ref{eq:B-L1}),~(\ref{eq:B-L2}),
\begin{equation}
{Y\over 2} = I^{3}_{\rm R} + {1\over 2} ({\rm B-L}) = -I^{3}_{\rm E} + {1\over 2}
({\rm B-L})~,
\end{equation}
where ${\rm B-L}$ is respectively given by Eqs.~(\ref{eq:B-L1}),~(\ref{eq:B-L3}), and
\begin{equation}
{Y\over 2} = -I^{3'}_{\rm R} + {1\over 2} ({\rm B-L}) = I^{3}_{\rm E} + {1\over 2}
({\rm B-L})~,
\end{equation}
where ${\rm B-L}$ is respectively given by Eqs.~(\ref{eq:B-L2}),~(\ref{eq:B-L3}).

In the SSB patterns of ${\rm E}_6$ via ${\rm SU}(3)^3$ there will therefore
often appear the intermediate symmetry group ${\rm SU}(3)_{\rm
C}\times {\rm SU}(2)_{\rm L}\times {\rm SU}(2)_{\rm (R)}\times {\rm
U}(1)_{\rm Y_L} \times {\rm U}(1)_{\rm Y_{(R)}}$. This intermediate
group can break down to ${\rm SU}(3)_{\rm C}\times {\rm SU}(2)_{\rm
L}\times {\rm SU}(2)_{\rm (R)}\times {\rm U}(1)_{\rm B-L}$ and the SSB
patterns which follow are just a generalisation of those written
previously for ${\rm SU}(3)_{\rm C}\times {\rm SU}(2)_{\rm L}\times
{\rm SU}(2)_{\rm R}\times {\rm U}(1)_{\rm B-L}$. 

The SSB of ${\rm SU}(3)_{\rm C}\times {\rm SU}(2)_{\rm
L}\times {\rm SU}(2)_{\rm (R)}\times {\rm U}(1)_{\rm Y_L} \times {\rm
U}(1)_{\rm Y_{(R)}}$ down to ${\rm G_{SM}}$ with associated defects
formation are given by
\begin{equation}
\label{eq:32211}
    {
3_{\rm C} ~2_{\rm L} ~2_{\rm (R)} ~1_{\rm Y_L} ~1_{\rm Y_{(R)}}
\left\{ 
\begin{array}{llllll} 
\overset{1}{\longrightarrow}&  3_{\rm C} ~2_{\rm L} ~2_{\rm (R)} ~1_{\rm B-L} 
&\left\{
\begin{array}{lllll} 
\overset{1}{\longrightarrow}& 3_{\rm C} ~2_{\rm L} ~1_{\rm (R)} ~1_{\rm B-L} &
\overset{2 ~(2)}{\longrightarrow} & {\rm G_{SM}}~(Z_2)\\
 \overset{2'~(2)}{\longrightarrow} & {\rm G_{SM}}~(Z_2)\\
 \end{array}
  \right.
\\
 \overset{1}{\longrightarrow}& 3_{\rm C} ~2_{\rm L} ~1_{\rm (R)} ~1_{\rm Y_{(R)}} ~1_{\rm Y_L} 
&\left\{
\begin{array}{lllll} 
\overset{2}{\longrightarrow}& 3_{\rm C} ~2_{\rm L} ~1_{\rm (R)} 
~1_{\rm B-L} & \overset{2 ~(2)}{\longrightarrow} &  {\rm G_{SM}} ~(Z_2)\\
 \overset{2 ~(2)}{\longrightarrow} & {\rm G_{SM}}~(Z_2)\\
 \end{array}
  \right.
\\
\overset{1,2}{\longrightarrow}& 3_{\rm C} ~2_{\rm L} ~1_{\rm (R)} ~1_{\rm B-L} 
& \overset{2 ~(2)}{\longrightarrow}  ~~~~{\rm G_{SM}}~(Z_2)\\
\overset{2'~(2)}{\longrightarrow}  & {\rm G_{SM}}~(Z_2)\\
  \end{array}
  \right.
    }
\end{equation}
We must count six times each SSB when we evaluate the number of
allowed schemes. 

We also have
  \begin{equation}
\label{eq:33L21}
\begin{array}{llllll} 
3_{\rm C} ~3_{\rm L} ~2_{\rm (R)} 1_{\rm (Y_R)} 
& \left\{ 
\begin{array}{clllllccccc}
\overset{1}{\longrightarrow}& 3_{\rm C} ~2_{\rm L} ~2_{\rm (R)} ~1_{\rm Y_{(R)}} ~1_{\rm Y_L}  &\overset{}{\longrightarrow}& \mbox{\rm Eq.~(\ref{eq:32211})} \\
 \overset{1}{\longrightarrow}& 3_{\rm C} ~2_{\rm L} ~1_{\rm (R)} ~1_{\rm Y_{(R)}} ~1_{\rm Y_L} &\overset{}{\longrightarrow}& \cdots \\
\overset{1}{\longrightarrow}& 3_{\rm C} ~2_{\rm L} ~2_{\rm (R)} ~1_{\rm B-L}  &\overset{}{\longrightarrow}& \cdots\\
\overset{1}{\longrightarrow}& 3_{\rm C} ~2_{\rm L} ~1_{\rm (R)} ~1_{\rm B-L} &\overset{2 ~(2)}{\longrightarrow}& {\rm G}_{\rm SM}~(Z_2)\\
\overset{2' ~(2)}{\longrightarrow}& {\rm G}_{\rm SM}~(Z_2)\\
  \end{array}
  \right.
\end{array}
\end{equation}
and
  \begin{equation}
\label{eq:3321}
\begin{array}{llllll} 
3_{\rm C} ~2_{\rm L} ~3_{\rm R} ~1_{\rm Y_L} &\left\{
    \begin{array}{lllll}
\overset{1}{\longrightarrow}& 3_{\rm C} ~2_{\rm (R)} ~2_{\rm L} ~1_{\rm Y_L} ~1_{\rm Y_{(R)}} &\overset{}{\longrightarrow}&  \mbox{\rm Eq.~(\ref{eq:32211})} \\
\overset{2'}{\longrightarrow}& 3_{\rm C} ~2_{\rm (R)}  ~2_{\rm L} ~1_{\rm B-L} &\overset{}{\longrightarrow}& \cdots\\
\overset{1}{\longrightarrow}& 3_{\rm C} ~2_{\rm L} ~1_{\rm (R)} ~1_{\rm Y_L} ~1_{\rm Y_{(R)}} &\overset{}{\longrightarrow}& \cdots\\
\overset{1}{\longrightarrow}& 3_{\rm C} ~2_{\rm L} ~1_{\rm (R)} ~1_{\rm B-L} &\overset{2 ~(2)}{\longrightarrow}& {\rm G}_{\rm SM}~(Z_2)\\
\overset{2' ~(2)}{\longrightarrow}& {\rm G}_{\rm SM}~(Z_2)\\
  \end{array}
  \right.
\end{array}
\end{equation}
also
 \begin{equation}
\label{eq:3311}
\begin{array}{llllll} 
3_{\rm C} ~3_{\rm L} ~1_{\rm (R)} ~1_{\rm Y_{(R)}}&\left\{ 
\begin{array}{clllll}
\overset{1}{\longrightarrow}& 3_{\rm C} ~2_{\rm L} ~1_{\rm Y_L} ~1_{\rm (R)} ~1_{\rm Y_{(R)}} &\overset{}{\longrightarrow} & \cdots\\
\overset{2'}{\longrightarrow}& 3_{\rm C} ~2_{\rm L} ~1_{\rm (R)} ~1_{\rm B-L} &
\overset{2 ~(2)}{\longrightarrow} & {\rm G_{SM}} ~(Z_2)\\
\overset{2' ~(2)}{\longrightarrow}&  {\rm G_{SM}} ~(Z_2)\\
  \end{array}
  \right.
\end{array}
\end{equation}

The SSB patterns of ${\rm E}_6$ via ${\rm SU}(3)_{\rm C}\times {\rm
SU}(3)_{\rm L}\times {\rm SU}(3)_{\rm R}$ with associated deftects
formation are given by
\begin{equation}
E_6
\overset{0}{\rightarrow} 3_{\rm C} ~3_{\rm L} ~3_{\rm R}       
  \left\{ 
    \begin{array}{cllllll}
\overset{1}{\longrightarrow}& 3_{\rm C} ~2_{\rm L} ~2_{\rm (R)} ~1_{\rm Y_{(R)}} ~1_{\rm Y_L} & \overset{}{\longrightarrow} & \mbox{\rm Eq.~(\ref{eq:32211})}\\
 \overset{1}{\longrightarrow} & 3_{\rm C} ~3_{\rm L} ~2_{\rm (R)} 1_{\rm (Y_R)} & \overset{}{\longrightarrow}&  \mbox{\rm Eq.~(\ref{eq:33L21})}
\\
\overset{1}{\longrightarrow}& 3_{\rm C} ~2_{\rm L} ~3_{\rm R} ~1_{\rm Y_L} & \overset{}{\longrightarrow} & \mbox{\rm Eq.~(\ref{eq:3321})}\\
\overset{1}{\longrightarrow}& 3_{\rm C} ~3_{\rm L} ~1_{\rm (R)} ~1_{\rm Y_{(R)}} & \overset{}{\longrightarrow} & \mbox{\rm Eq.~(\ref{eq:3311})}\\
\overset{1}{\longrightarrow}& 3_{\rm C} ~2_{\rm L} ~1_{\rm (R)} ~1_{\rm Y_{(R)}} ~1_{\rm Y_L}& \overset{}{\longrightarrow} & \cdots\\
\overset{1}{\longrightarrow}& 3_{\rm C} ~2_{\rm L} ~2_{\rm (R)} ~1_{\rm B-L} 
& \overset{}{\longrightarrow} & \cdots\\
\overset{1}{\longrightarrow}& 3_{\rm C} ~2_{\rm L} ~1_{\rm (R)} ~1_{\rm B-L} 
& \overset{2 ~(2)}{\longrightarrow}  & {\rm G_{SM}} ~(Z_2)\\
\overset{1 ~(1,2)}{\longrightarrow}& {\rm G}_{\rm SM}~(Z_2)\\
  \end{array}
  \right.
\end{equation}
There are more direct breakings which are given by
\begin{equation}
    {
E_6\left\{ \begin{array}{clllllcccc} \overset{1}{\longrightarrow}& 3_{\rm C} ~2_{\rm L} ~2_{\rm (R)} ~1_{\rm Y_L} ~1_{\rm Y_{(R)}} &\overset{}{\longrightarrow}& \mbox{\rm Eq.~(\ref{eq:32211})}\\
\overset{1}{\longrightarrow}& 3_{\rm C} ~3_{\rm L} ~2_{\rm (R)} ~1_{\rm Y_{(R)}} &\overset{}{\longrightarrow}& \mbox{\rm Eq.~(\ref{eq:33L21})}  \\
\overset{1}{\longrightarrow}& 3_{\rm C} ~2_{\rm L} ~3_{\rm R} ~1_{\rm Y_L} &\overset{}{\longrightarrow}& \mbox{\rm Eq.~(\ref{eq:3321})} \\
\overset{1}{\longrightarrow}& 3_{\rm C} ~2_{\rm L} ~1_{\rm (R)} ~1_{\rm Y_L} ~1_{\rm Y_{(R)}} &\overset{}{\longrightarrow}& \cdots\\
\overset{1}{\longrightarrow}& 3_{\rm C} ~2_{\rm L} ~2_{\rm (R)} ~1_{\rm B-L} &\overset{}{\longrightarrow}& \cdots\\ 
\overset{1}{\longrightarrow}& 3_{\rm C} ~2_{\rm L} ~1_{\rm (R)} ~1_{\rm B-L} &\overset{2 ~(2)}{\longrightarrow}& {\rm G}_{\rm SM}~(Z_2)\\ 
  \end{array}
  \right.
    }
\end{equation}

There is now the possibility of having inflation and embedded strings
forming at the end of inflation together with R-parity conservation.
The total number of schemes satisfying cosmological constraints for
defects is 1086, with 522 schemes leading to the formation of
topological strings at the end of inflation with conserved R-parity,
384 with topological strings and broken R-parity, 162 models with
embedded strings and broken R-parity and 18 schemes with embedded
strings and conserved R-parity. When the constraint of leptogenesis is
added, we find 444 schemes leading to the formation of topological
strings at the end of inflation with conserved R-parity, 312 with
topological strings and broken R-parity and 138 models with embedded
strings and broken R-parity. There is not any model with embedded
strings and R-parity. In conclusion, there are 444 models
satisfying all constraints, and they all lead to the formation of
topological cosmic strings at the end of inflation.

\subsubsection{Breaking ${\rm E}_6$ via ${\rm SU}(6) \times {\rm SU}(2)$}

We end with ${\rm E}_6 \supset {\rm SU}(6) \times {\rm SU}(2)$. There
are two possibilities, namely ${\rm SU}(6)\times {\rm
SU}(2)_{\rm L}$ or ${\rm SU}(6)_{}\times {\rm SU}(2)_{\rm R}$. A third
possibility would be ${\rm SU}(6) \times {\rm SU}(2)_{\rm I}$ where
${\rm SU}(2)_{\rm I}$ is called {\sl inert}~\cite{londonrosner}
because it is orthogonal to ${\rm G_{SM}}$ which is embedded
completely inside ${\rm SU}(6)$. However, since in SUSY models
this embedding is not compatible with the proton lifetime, we
do not study it.

We first study ${\rm E}_6 \supset {\rm SU}(6) \times {\rm SU}(2)_{\rm
L}$. We consider the following SSB patterns of ${\rm E}_6$ via ${\rm
SU}(6) \times {\rm SU}(2)_{\rm L}$ which we write together with
the more direct breakings:
\begin{equation}
\label{eq:62L}  
\begin{array}{lll}
\begin{array}{ccc}
{\rm E}_6  &\overset{0}{\longrightarrow}  & 6 ~2_{\rm L} \\ \\
&\mbox{or}
&{\rm E}_6
\end{array}
&\left\{ 
    \begin{array}{cllll}
\overset{1}{\longrightarrow}& 3_{\rm C} ~3_{\rm R} ~2_{\rm L} ~1_{\rm Y_L} 
& \overset{}{\longrightarrow}  & \mbox{\rm Eq. (\ref{eq:3321})} \\
\overset{1}{\longrightarrow}& 4_{\rm C} ~2_{\rm L} ~2_{\rm R} ~1_{\rm V'} 
&\overset{}{\longrightarrow} & \mbox{\rm Eq. (\ref{eq:ps6})}\\
\overset{1}{\longrightarrow}& 4_{\rm C'} ~2_{\rm L} ~2_{\rm G} ~1_{\rm V'} 
&\overset{}{\longrightarrow} & \mbox{\rm Eq. (\ref{eq:psp6})}
\\
\overset{0}{\longrightarrow}& 4_{\rm C} ~2_{\rm L} ~2_{\rm R} &\overset{}{\longrightarrow}& \cdots\\
\overset{1}{\longrightarrow}& 4_{\rm C} ~2_{\rm L} ~1_{\rm R} ~1_{\rm V'} &\overset{}{\longrightarrow}& \cdots\\
\overset{1}{\longrightarrow}& 4_{\rm C} ~2_{\rm L}~1_{\rm R} &\overset{}{\longrightarrow}& \cdots\\
\overset{1}{\longrightarrow}& 3_{\rm C} ~2_{\rm L} ~2_{\rm (R)} ~1_{\rm Y_L} ~1_{\rm Y_{\rm (R)}} &\overset{}{\longrightarrow}& \mbox{\rm Eq. (\ref{eq:psp6})}\\
\overset{1}{\longrightarrow}& 3_{\rm C} ~2_{\rm L} ~2_{\rm (R)} ~1_{\rm B-L} &\overset{}{\longrightarrow}& \cdots\\
\overset{1}{\longrightarrow}& 3_{\rm C} ~2_{\rm L} ~1_{\rm (R)} ~1_{\rm Y_L} ~1_{\rm Y_{(R)}} &\overset{}{\longrightarrow}& \cdots\\
\overset{1}{\longrightarrow}& 3_{\rm C} ~2_{\rm L} ~1_{\rm (R)} ~1_{\rm B-L} &\overset{2 ~(2)}{\longrightarrow}& {\rm G}_{\rm SM}~(Z_2)\\
\overset{1 ~(1,2)}{\longrightarrow}& {\rm G}_{\rm SM}~(Z_2) \\ 
\end{array}
  \right.\\
\end{array}
\end{equation}

The ${\rm SU}(2)_{\rm L}$ of ${\rm G_{SM}}$ can also be contained in
SU(6), so that ${\rm E}_6$ breaks down to ${\rm SU}(6) \times {\rm
SU(2)_R}$. We consider the following SSB schemes of ${\rm E}_6$ via
${\rm SU}(6) \times {\rm SU}(2)_{\rm R}$ which we write
together with the more direct breakings:
\begin{equation} 
\begin{array}{clllll}
\begin{array}{ccc}
{\rm E}_6  &\overset{0}{\longrightarrow}  & 6 ~2_{\rm R} \\ \\
&\mbox{or}
&{\rm E}_6 
\end{array}
& {
\left\{ 
    \begin{array}{clllll}
 \overset{1}{\longrightarrow}& 4_{\rm C} ~2_{\rm L} ~2_{\rm R} ~1_{V'}
 & \overset{}{\longrightarrow} ~~~~\mbox{\rm Eq. (\ref{eq:ps6})} \\ 
\overset{1}{\longrightarrow}& 4_{\rm C} ~2_{\rm L} ~1_{\rm R} ~1_{\rm
 V'} & \overset{}{\longrightarrow}~~~~ \cdots\\
 \overset{0}{\longrightarrow}& 4_{\rm C} ~2_{\rm L} ~2_{\rm R} &
 \overset{}{\longrightarrow} ~~~~ \cdots \\
\overset{1}{\longrightarrow}& 4_{\rm C} ~2_{\rm L} ~1_{\rm R} &
 \overset{}{\longrightarrow}~~~~ \cdots\\
  \overset{1}{\longrightarrow}& 3_{\rm C} ~2_{\rm L} ~2_{\rm R} ~1_{\rm
 B-L} ~1_{\rm V'} & \overset{}{\longrightarrow}~~~~ \cdots\\
 \overset{1}{\longrightarrow}& 3_{\rm C} ~2_{\rm L} ~2_{\rm R} ~1_{\rm
 B-L} & \overset{}{\longrightarrow}~~~~ \cdots\\
 \overset{1}{\longrightarrow}& 3_{\rm C} ~2_{\rm L} ~1_{\rm R} ~1_{\rm
 B-L} & \overset{2 (2)}{\longrightarrow}~~~~ {\rm G}_{\rm SM}~(Z_2)\\
 \overset{1 ~(1,2)}{\longrightarrow}& {\rm G}_{\rm SM}~(Z_2) \\
\end{array}
  \right.
 }\\  \end{array}
\end{equation}

The total number of schemes satisfying cosmological constraints for
defects is 1270, with 664 schemes leading to the formation of
topological strings at the end of inflation with conserved R-parity,
422 with topological strings and broken R-parity, 170 models with
embedded strings and broken R-parity and 12 schemes with embedded
strings and conserved R-parity. When the constraint of leptogenesis is
added, we find 534 models satisfying all constraints, and they all
lead to the formation of topological cosmic strings at the end of
inflation.

\subsection{SU(7)}

SU(7) is the second group of rank 6. The embeddings which one can
choose are
$${\rm SU}(7)\supset {\rm SU}(6)_{\rm SM}\times {\rm U}(1) ~~\mbox{with}~~
{\rm SU}(6)_{\rm SM} \supset {\rm G_{\rm SM}}, $$ 
$${\rm SU(7)}\supset {\rm SU}(6)\times
{\rm U}(1)~~\mbox{with}~~ {\rm SU}(6) \times {\rm U}(1) \supset {\rm G}_{\rm SM},$$ 
$${\rm SU}(7)\supset {\rm SU}(4)_{\rm C}\times {\rm SU}(3)_{\rm L}\times {\rm U}(1),$$$${\rm SU}(7)\supset {\rm SU}(5)_{\rm C} \times {\rm SU}(2)_{\rm L}
\times {\rm U}(1) ~~\mbox{with}~~ {\rm SU}(5)_{\rm C} \times {\rm
SU}(2)_{\rm L} \times {\rm U}(1) \supset {\rm G_{\rm SM}} ,$$
$${\rm SU}(7)\supset {\rm SU}(5)_{\rm SM} \times {\rm SU}(2) \times
{\rm U}(1) ~~\mbox{with}~~ {\rm SU}(5)_{\rm SM} \supset {\rm G_{\rm SM}} .$$
In the first and later cases ${\rm G_{\rm SM}}$ is completely embedded
in the ${\rm SU}(6)$ (${\rm SU}(5)$) factor.

The only possibility for getting inflation without defect formation at
the end or after, is if we have the later scheme, where the SU(2)
factor is orthogonal to ${\rm G_{\rm SM}}$. The SSB patterns which
could accommodate an epoch of inflation with no defect (of any kind)
formation at a later stage are
\begin{equation}
{\rm SU}(7)\overset{1}{\rightarrow} 5_{\rm SM} ~2 ~1
\left\{ \begin{array}{cclcc}
\overset{2}{\rightarrow}& 
 5_{\rm SM} ~2 &\overset{1}{\rightarrow} {\rm G}_{\rm SM} ~2 \overset{0}{\rightarrow} {\rm G}_{\rm SM}
\\
\overset{1}{\rightarrow}& 
 {\rm G}_{\rm SM} ~2 ~1&\overset{2}{\rightarrow} {\rm G}_{\rm SM} ~2 \overset{0}{\rightarrow} {\rm G}_{\rm SM}
\\
\overset{1,~2}{\rightarrow}& 
 {\rm G}_{\rm SM} ~2 &\overset{0}{\rightarrow} {\rm G}_{\rm SM}
\end{array}
  \right.
\end{equation}
However, these models are inconsistent with proton lifetime
measurements and minimal SU(7) does not predict neutrino masses. These
models are therefore incompatible with high energy physics
phenomenology.

\subsection{Higher Rank Groups}

There are two groups of rank 7, namely SO(14) and ${\rm SU}(8)$. These
groups are particularly interesting since they both contain ${\rm
U}(1)_{\rm B-L}$. In what follows, we discuss the embeddings of the
${\rm G_{SM}}$ in these groups and we then comment on the SSB
patterns, without writing down explicitely all of them. We just aim to
extract those scenarios which can lead to inflation without cosmic string
formation at the end of inflation or afterwards.

SO(14) has only two maximal sub-algebras, 
$${\rm SU}(7)\times {\rm U}(1),$$ 
$$ {\rm SO}(10) \times {\rm SU}(2)\times {\rm SU}(2)~.$$ The only
possibility for getting inflation without strings in the first case,
is to embed the standard model in ${\rm SU}(5)_{\rm SM} \subset {\rm
SU}(7)$ in ${\rm SU}(5)_{\rm SM} \times {\rm SU}(2) \times {\rm U}(1)
\subset {\rm SU}(7)$ so that the SU(2) and the two U(1)s in ${\rm
SU}(5)_{\rm SM} \times {\rm SU}(2) \times {\rm U}(1) \times {\rm U}(1)
\subset {\rm SO}(14)$ are orthogonal to ${\rm G_{SM}}$. These models
are also inconsistent with observations from both particle physics
and cosmological point of view.

If we consider the maximal sub-algebra ${\rm SO}(10) \times {\rm
SU}(2)\times {\rm SU}(2)$, then the only way would be to embed
${\rm G_{SM}}$ in SO(10) so that the two SU(2) factors are inert, and
break SU(2) down to identity after the breaking of SO(10). These
models are also inconsistent with observations.

SU(8) maximal sub-groups are ${\rm SU}(7) \times {\rm U}(1)$ and
${\rm SU}(m) \times {\rm SU}(n) \times {\rm U}(1)$ where $m+n =
8$. One may have ${\rm SU}(m) \supset {\rm SU}(3)_{\rm C}$ and ${\rm
SU}(m) \supset {\rm SU}(2)_{\rm L}$ or for $m ~(n) \geq 5$ embed ${\rm
G_{SM}}$ in ${\rm SU}(m)$ so that ${\rm SU}(n) \times {\rm U}(1)$ is
orthogonal to it. The only way to get inflation without strings in the
first case is to embed ${\rm G_{SM}}$ entirely in SU(7), break ${\rm
U}(1)$ before inflation and we are left with the SU(7) cases
mentioned above. These models are inconsistent from both the
cosmological and particle physics requirements that we have. One can
easily show that in the second case where ${\rm SU}(m,n) \supset {\rm
SU}(3)_{\rm C}$ and ${\rm SU}(n,m) \supset {\rm SU}(2)_{\rm L}$ all
SSB patterns with inflation and leptogenesis lead to the formations of
cosmic strings at the end of inflation (topological or embedded ones)
and if unbroken R-parity is required, the strings are topological and
topologically stable down to low energies. The only possibility for
having inflation without strings might be the last case where $m$ or
$n > 5$ and to embed ${\rm G_{SM}}$ in ${\rm SU}(m,n)$. But here
again, it seems impossible to fit leptogenesis after inflation.
Therefore all SU(8) models with standard hybrid inflation and
baryogenesis lead to the formation of cosmic strings at the end of
inflation.

Finally, there is one group of rank 8, SU(9). Following the same
procedure as for the SU(8) case, we conclude that none of the SSB schemes
lead to inflation without strings after the end of the inflationary era.

\section{Conclusion and Discussion}

 Current data from the realm of cosmology strongly support an
early inflationary era. In addition, current CMB temperature
anisotropies data minimize a possible contribution from cosmic
strings. On the other hand, many GUTs naturally lead to cosmic string
formation. We are thus faced with a crucial quest, namely how often
GUTs lead to cosmic string formation? Or, in other words, which is a
natural inflationary scenario? Answering these questions is the
motivation of our study.

In the context of SUSY GUTs, we have studied the cosmological
implications of SSB patterns from grand unified gauge groups ${\rm
  G_{GUT}}$ down to the standard model gauge group ${\rm G_{SM}}$. The
aim is to select all the schemes which can satisfy both cosmological
and particle physics constraints, among them: lead to inflation and
solve the GUT monopole problem, explain the baryon asymmetry of the
universe, predict neutrino masses and lead to automatic R-parity
conservation.  To perform this analysis, we limit ourselves to simple
gauge groups which contain ${\rm G_{SM}}$, have a complex
representation, are anomaly free, and have a rank not greater than 8.
Such gauge groups are: SU(5), SO(10), SU(6), E$_6$, SU(7), SO(14),
SU(8), and SU(9).  We take a large number of possible embeddings of
${\rm G_{SM}}$ in ${\rm G_{GUT}}$ and we list in detail all possible
SSB patterns of ${\rm G_{GUT}}$ down to ${\rm G_{SM}}$.  We also
investigate whether defects are formed during the SSB schemes and of
which kind they are.  We assume standard hybrid inflation which
emerges naturally in SUSY GUTs, the inflaton field being a linear
combination of a singlet field and one component of the complex GUT
Higgs fields which are used to lower the rank of the group by one
unit. We then examine whether monopoles or domain walls are formed
after the end of inflation. We disregard such SSB patterns.  We also
disregard SSB schemes with broken R-parity.  To be consistent with
leptogenesis, we require that the gauged ${\rm B-L}$ symmetry, which
is contained in GUTs which predict neutrino masses via the see-saw
mechanism and unbroken R-parity, is broken at the end inflation. This,
for example, implies that we throw away SU(6) or SU(7). We then
compare the SSB patterns where topological cosmic strings or embedded
strings are formed after inflation with respect to the SSB patterns
where there are no defects at all after the end of the inflationary
era.

Among the SSB schemes which are compatible with high energy physics
and cosmology, we did not find any without strings after inflation.
One should thus only consider mixed models, where inflation co-exists
with cosmic strings.  On the other hand, various cosmological issues,
and in particular the CMB temperature anisotropies, set bounds to the
cosmic string contribution. This can help constraining or ruling out
realistic GUT models where the string contribution can always be
computed. One may also have to reconsider the validity of the whole
theoretical framework. 

We also find the existence of SSB patterns, for GUTs based on gauge
groups which have rank greater than six, which predict the formation of
secondary string networks at lower energies. Finally, in all models
with strings and inflation, the strings forming at the end of
inflation are the so-callled B-L cosmic strings \cite{leptrj}. Their
contribution to the baryon assymmetry of the universe is non-negligible 
and may compete with the non-thermal process of
leptogenesis from reheating.

\appendix
\section{Maximal Sub-groups}
\label{subalgebras}

We list the maximal sub-groups of each GUT which is
studied in this paper. They are necessary for finding the SSB
patterns. Most of the information given here is taken from
Ref.~\cite{slansky}. We only consider maximal regular sub-groups
because it is very difficult and unatural to get ${\rm G_{SM}}$ via
maximal special sub-groups. As discussed in Sec.~\ref{sec-discrete} some
discrete symmetries may also appear during the SSB patterns, they do
not appear here. Sometimes, there is more than one possibilities to
embbed a maximal regular sub-group in the GUT, and we add indices to
make this explicit to the reader. In general, a subscript C means
that this groups contains ${\rm SU}(3)_{\rm C}$ as a sub-group and
subscript L means that this groups contains ${\rm SU}(2)_{\rm L}$ as a
sub-group. Definitions of indices for each maximal sub-group will be
obvious to the reader in each section dedicated to a given GUT.
\begin{table}[hhh]
\begin{center}
\begin{tabular}{c|c|c}
Rank & Group & Maximal sub-algebras \\ \hline
4 & ${\rm SU}(5)$ & ${\rm SU}(3)_{\rm C} \times {\rm SU}(2)_{\rm L} \times  {\rm U}(1)_{\rm Y}$  \\
  &         & ${\rm SU}(4) \times {\rm U}(1) $ \\ \hline
5  & ${\rm SO}(10)$& ${\rm SU}(5) \times {\rm U}(1)_{\rm V}$ \\
  &         & ${\rm SU}(5)_{\rm F} \times {\rm U}(1)_{\rm V}$ \\
  &         & ${\rm SU}(4)_{\rm C} \times {\rm SU}(2)_{\rm L} \times {\rm SU}(2)_{\rm R} = {\rm G_{PS}} $ \\ \hline
  & ${\rm SU}(6)$ & ${\rm SU}(5) \times {\rm U}(1)_{\rm 6}$ \\ 
  &         & $ {\rm SU}(4)_{\rm C} \times {\rm SU}(2)_{\rm L} \times {\rm U}(1) $ \\ 
  &         & ${\rm SU}(3)_{\rm C} \times {\rm SU}(3)_{\rm L} \times {\rm U}(1) $ \\ \hline
6 & ${\rm E}_6$   & ${\rm SO}(10) \times {\rm U}(1)_{\rm V'}$ \\ 
  &         & ${\rm SU}(6) \times {\rm SU}(2)_{\rm R} $ \\ 
  &         & ${\rm SU}(6) \times {\rm SU}(2)_{\rm L}$ \\ 
  &         & ${\rm SU}(3)_{\rm C} \times {\rm SU}(3)_{\rm L} \times {\rm SU}(3)_{\rm (R)}$ \\ \hline
  & ${\rm SU}(7)$ & ${\rm SU}(6)_{\rm SM}  \times {\rm U}(1)$ \\ 
  &         & ${\rm SU}(5)_{\rm C} \times {\rm SU}(2)_{\rm L} \times  \times {\rm U}(1) $ \\ 
  &         & ${\rm SU}(5)_{\rm SM} \times {\rm SU}(2) \times {\rm U}(1) $ \\ 
  &         & ${\rm SU}(4)_{\rm C} \times {\rm SU}(3)_{\rm L} \times  {\rm U}(1) $ \\ \hline
7 & ${\rm SO}(14)$& ${\rm SU}(7) \times {\rm U}(1)$ \\  
  &         & ${\rm SO}(10) \times {\rm SU}(2) \times {\rm SU}(2) $ \\ \hline
  & ${\rm SU}(8)$ & ${\rm SU}(7) \times {\rm U}(1)$ \\ 
  &         & ${\rm SU}(6) \times {\rm SU}(2) \times  {\rm U}(1) $ \\
  &         & ${\rm SU}(5) \times {\rm SU}(3) \times  {\rm U}(1) $ \\ 
  &         & ${\rm SU}(4) \times {\rm SU}(4) \times {\rm U}(1) $ \\ \hline
8 & ${\rm SU}(9)$ & ${\rm SU}(8) \times {\rm U}(1)$ \\
  &         & ${\rm SU}(7) \times {\rm SU}(2) \times  {\rm U}(1)$ \\
  &         & ${\rm SU}(6) \times {\rm SU}(3)  \times  {\rm U}(1)$ \\
  &         & ${\rm SU}(5) \times {\rm SU}(4) \times  {\rm U}(1) $ \\
\end{tabular}
\caption{Maximal regular sub-groups of grand unification gauge groups
with rank not greater than 8.}
\end{center}\label{sub-group1}
\end{table}

\section*{Acknowledgments}

We would like to thank U. Ellwanger, I.~Gogoladze, A.~Perez-Lorenzana,
P.~Peter, M.~Postma, G. Senjanovi\`c and C.-M.~Viallet for useful
discussions. The work of R.J. and J.R. was partially supported by the
EC network HPRN-CT-2000-00152.  M.S. acknoweledges financial support
from E.L.K.E. ({\sl Special Account for Reasearch}), University of
Athens, Hellas.


\begin{thebibliography}{50}

\bibitem{SK} Y.~Fukuda {\it et al.}  [Super-Kamiokande Collaboration],
Phys.\ Rev.\ Lett.\  {\bf 81}, 1562 (1998).

\bibitem{SNO} Q.~R.~Ahmad {\it et al.}  [SNO Collaboration], Phys.\
Rev.\ Lett.\ {\bf 87}, 071301 (2001).

\bibitem{kamland} K. Eguchi {\it et al.} [KamLAND Collaboration], Phys.\
Rev.\ Lett.\ {\bf 90}, 021802 (2003).


\bibitem{see-saw} M.\ Gell-Mann, P. \ Ramond and R.\ Slansky, in :
Supergravity, eds. P.~ van Nieuwenhuizen and D.~ Freeman (North Holland,
Amsterdam, 1979) p.~315; T. Yanagida, Prog.\ Th.\ Phys.\ B {\bf 135}, 66 (1979);
R.~N.~ Mohapatra and G.~Senjanovic, \emph{Neutrino mass and spontaneous parity
  violation},  Phys.\ Rev.\ Lett.\ {\bf 44}, 912 (1980).

\bibitem{PDG} K.\ Hogiwara {\em et al.}, Phys. Rev. D{\bf 66}, 010001 (2002).

\bibitem{kibble} T.\ W.\ B.\ Kibble, J.\ Phys.\ A{\bf 9}, 387 (1976).

\bibitem{wmap} G.\ Hinshaw {\em et al.} ``First Year Wilkinson Microwave 
Anisotropy Probe (WMAP) Observations:  Angular Power Spectrum'',{\tt 
astro-ph/0302217}.

\bibitem{bouchet}  
F.\ R.\ Bouchet, P.\ Peter, A.\ Riazuelo, and M.\ Sakellariadou,
Phys.\ Rev. D{\bf 65}, 021301 (2002).

\bibitem{textures} N.~Turok, Phys.\ Rev.\ Lett.\ {\bf 63}, 2625 (1989).

\bibitem{ShelVil} A. Vilenkin and E.P.S. Shellard, {\em ``Cosmic strings
and other topological defects''}, Cambridge monographs on mathematical
physics, Cambridge University Press, England, 1994.

\bibitem{hk} M.\ B.\ Hindmarsh and T.\ W.\ B.\ Kibble, Rept.\ Prog.\ Phys.\
{\bf 58}, 477 (1995).

\bibitem{emb1} T.~Vachaspati and M.~Barriola, Phys.\ Rev.\ Lett.\ {\bf
69}, 1867 (1992).

\bibitem{emb2} T.~Vachaspati, Phys.\ Rev.\ Lett.\ {\bf 78}, 1977 (1992);
Nucl.\ Phys.\ {\bf B397}, 648 (1993).

\bibitem{periv} M.~James, L.~Perivolaropoulos, and T.~Vachaspati,
Phys.\ Rev. D{\bf 46}, 5232 (1992); Nucl.\ Phys.\ {\bf B395}, 534
(1993); M.\ Goodband and M. Hindmarsh, Phys.\ Lett.\ {\bf B363}, 58 (1995).

\bibitem{stab1} M.~Nagasawa and R.~Brandenberger, Phys.\ Rev.\ D{\bf 67}, 043504 (2003).

\bibitem{stab2} T.~Vachaspati and R.~Watkins, Phys.\ Lett.\ B{\bf 318}
163 (1993).

\bibitem{unst} M.~Barriola, T.~Vachaspati, and M.~Bucher, Phys.\ Rev.\ 
D{\bf 50} 2819 (1994).

\bibitem{bnc} R.~A.~Brandt and F.~Neri, Nucl.\ Phys.\ {\bf B161}, 253
  (1979); S.~Coleman, Erice Lectures (1981).

\bibitem{calzsakel} E.\ Calzetta and M.\ Sakellariadou, Phys.\ Rev.\ D{\bf 45},
2802 (1992); E.\ Calzetta and M.\ Sakellariadou, Phys.\ Rev.\ D{\bf 47},
3184 (1993).

\bibitem{sakel} M.\ Sakellariadou, ({\tt astro-ph/9911497}), in
       ``Recent Developments in Gravitation'', 
             Proceedings of the Spanish Relativity Meeting, ERE-99, Bilbao
       (Spain), 7-10 Septembre 1999, J.\ Ibanez, ed., 113 (2000).

\bibitem{LythRiotto} D.\ H.\ Lyth and A.\ Riotto, Phys.\ Rept.\ {\bf 314},~1 (1999) 

\bibitem{Linde} A.\ D.\ Linde, Phys.\ Rev.\ D{\bf 49}, 748 (1994). 

\bibitem{Cop} E.\ J. Copeland, A.\ R.\ Liddle, D.\ H.\ Lyth, E.\ D.\ Stewart and
D.\ Wands, Phys.\ Rev.\ D{\bf 49}, 6410 (1994).

\bibitem{Dvasha} G.\ Dvali, Q.\ Shafi and R.\ Schaefer,
Phys.\ Rev.\ Lett.\ {\bf  73}, 1886 (1994).

\bibitem{Lindechaotic} A.\ D.\ Linde, Phys.\ Lett.\ B {\bf 129}, 177 (1983). 

\bibitem{buchmullercovi} W.\ Buchmuller, L.\ Covi and D.\ Delepine,
Phys.\ Lett.\ B{\bf 491}, 183 (2000).

\bibitem{SO(10)} A-C. Davis and R.\ Jeannerot, Phys.\ Rev.\ D{\bf 52}, 7220 (1995).

\bibitem{rjinflsusy} R.\ Jeannerot, Phys.\ Rev.\ D{\bf 56}, 6205 (1997). 

\bibitem{acdkss}  
  B.~Allen, R.R.~Caldwell, S.~Dodelson, L.~Knox, E.P.S.~Shellard and
  A.~Stebbins, \prl {\bf 79}, 2624 (1997).

\bibitem{mark}
  C.~Contaldi, M.~Hindmarsh and J.~Magueijo, Phys.\ Rev.\ Lett.\ {\bf 82}, 679
  (1999).

\bibitem{ringeval} A.~Riazuelo and C.\ Ringeval, {\sl private communication} (2003).

\bibitem{vhs}  G.\ R.\ Vincent, M.\ Hindmarsh and M.\ Sakellariadou,
Phys.\ Rev.\ D{\bf 56}, 637 (1997).

\bibitem{wiggles} 
  L.~Pogosian and T.~Vachaspati, Phys.\ Rev.\ D{\bf 60}, 083504 (1999).
  
\bibitem{sakel2} L.\ Perivoralopoulos, Phys.\ Rev.\ D{\bf 48}, 1530
  (1993); M.\ Sakellariadou, Int.\ J.\ of Theor.\ Phys.\, {\bf 36},
  No. 11, 847 (1997); A.\ Gangui, L.\ Pogosian and S.\ Winitzk, Phys.\ 
  Rev.\ D{\bf 64}, 043001 (2001).

\bibitem{d-term} 
W.\ Fishler, H.\ P.\ Nilles, J.\ Polchinski, S.\ Raby, and L.\
Susskind, Phys.\ Rev.\ Lett. {\bf 47}, 757 (1981).

\bibitem{fi}
P.\ Fayet, and J.\ Iliopoulos, Phys.\ Lett.\ {\bf 51B}, 461 (1974).


\bibitem{bindva} P.\ Binetruy and  G.\ R.\ Dvali, Phys.\ Lett.\ B388, 241 (1996).

\bibitem{shifted1} R.\ Jeannerot, S.\ Khalil, G.\ Lazarides and
Q. Shafi, JHEP {\bf 0010}, 012 (2000 ).

\bibitem{shifted} R.\ Jeannerot, S.\ Khalil and G.\ Lazarides, JHEP
{\bf 0207}, 069 (2002).

\bibitem{monopb} R.\ Jeannerot, S.\ Khalil and G.\ Lazarides, Proceedings
of Cairo International Conference on High-Energy Physics (CICHEP
2001), Cairo, Egypt, 9-14 Jan 2001.

\bibitem{sarkar}
J.\ A.\ Adams and S.~Sarkar, Nucl.\ Phys.\ B{\bf 503}, 405 (1997).

\bibitem{tetradismairi}
N.~Tetradis and M.~Sakellariadou, {\tt hep-ph/9806461} (1998).

\bibitem{langacker} P.\ Langacker, Phys.\ Rept.\ {\bf 72}, 185 (1981).

\bibitem{ross} G.\ G.\ Ross, ``Grand Unified Theories'', Reading, Usa:
Benjamin/cummings (1984) 497 P. ( Frontiers In Physics, 60).

\bibitem{slansky}
R.\ Slansky, Phys.\ Rep.\ {\bf 79},1 (1981).

\bibitem{cmm} M.\ Dubois-Violette, M.\ Talon and C.\ M.\ Viallet,
Com.\ in Math.\ Phys. {\bf 102}, 105 (1985); M.\ Dubois-Violette, M.\
Talon and C.\ M.\ Viallet, Annales de l'Inst.\ Henri Poincar\'e {\bf
44}, 103 (1986).


\bibitem{Yanagida} M. Fukugita and T. Yanagida, Phys. Lett.B 174:45
(1986).

\bibitem{DineKus} M. Dine and A. Kusenko, Submitted to Rev.Mod.Phys.,
hep-ph/0303065.

\bibitem{BP} W. Buchm\"uller and M. Plumacher, Phys. Rept 320, 329,
(1999) and Refs therein.

\bibitem{Rubakov} V.A. Kuzmin, V.A. Rubakov and M.E. Shaposhnikov,
Phys. Lett. B 155, 36 (1985).

\bibitem{Lazsha} G. Lazarides, Robert K. Schaefer and Q. Shafi,
Phys.Rev.D56, 1324 (1997); G. Lazarides, Q. Shafi, N.D. Vlachos,
Phys. Lett. B427, 53 (1998).

\bibitem{leptrj} R. Jeannerot, Phys. Rev. Lett. 77, 3292 (1996).

\bibitem{leptrj2} R. Jeannerot, Proceedings de COSMO-97, International
Workshop on {\em Particle Physics and the Early Universe}, 15-19
September 1997, Ambleside, Lake district, England.


\bibitem{hw} P. Horava and E. Witten, Nucl.Phys. B460 (1996) 506.

\bibitem{quevedo} F.~Quevedo, Nucl.\ Phys.\ Proc.\ Suppl.\ {\bf 62}, 134 (1998). 

\bibitem{ub} P.~ Candelas, E.~ Perevalov and G.~ Rajesh, 
Nucl.\ Phys.\ B{\bf 507} 445 (1997).

\bibitem{Martin} S.\ P.\ Martin, Phys.\ Rev.\ D{\bf 46}, 2769 (1992 ).

\bibitem{ibanez}
L. E.\ Ibanez, G. G.\ Ross, Nucl.\ Phys.\ {\bf B368}, 3 (1992).

\bibitem{montigny}M.\ de\ Montigny, M.\ Masip, Phys.\ Rev. D{\bf 49},
3734 (1994).

\bibitem{Kibblelaz} T.\ W.\ B.\ Kibble, G.\ Lazarides and Q.\ Shafi,
Phys.\ Rev.\ D {\bf 26}, 435 (1982).

\bibitem{moha2} D.\ Chang, R.\ N.\ Mohapatra and M.\ K.\ Parida,
Phys.\ Rev.\ Lett.\ {\bf 52}, 1072 (1984).

\bibitem{masiero} L.\ Covi, G.\ Mangano, A.\ Masiero, G.\ Miele,
Phys.\ Lett.\ B {\bf 424}, 25 (1998).

\bibitem{harada} J.~Harada, Submitted to JHEP, hep-ph/0305015.

\bibitem{georgiglashow} H.~Georgi and S.L.~Glashow, Phys.\ Rev.\ Lett. {\bf 32}, 438 (1974).

\bibitem{flippedsu5} S.~M.~Barr, Phys.\ Lett.\ B{\bf 112},219 (1982); J.~P.~Derendinger, 
J.~E.~Kim and D.~V.~Nanopoulos, Phys.\ Lett.\ B{\bf 139},170 (1984); 
I.~Antoniadis, J.~Ellis, J.~S.~Hagelin and D.~V.~Nanopoulos, Phys.\ Lett.\ B{\bf 194},231 (1987); 
J.~Ellis, J.~L.~Lopez and D.~V.~Nanopoulos, Phys.\ Lett.\ B{\bf 371},65 (1996).


\bibitem{sato} J.\ Sato, Prog.\ Theor.\ Phys.\ {\bf 96}, 597 (1996).


\bibitem{londonrosner} D.\ London and J.\ Rosner, Phys.\ Rev.\ D {\bf 34}, 435 (1986).
\end{thebibliography}
\end{document}